\documentclass{jfm-draft}
\usepackage{bm}
\usepackage{natbib}
\usepackage{epsfig}
\usepackage{amssymb}
\usepackage[dvips]{color}
\usepackage{lscape}
\usepackage{longtable}
\usepackage{rotating}
\usepackage{amsmath}
\newcommand{\be}{\begin{equation}}
\newcommand{\ee}{\end{equation}}
\newcommand{\ba}{\begin{eqnarray}}
\newcommand{\ea}{\end{eqnarray}}

\newcommand{\bigT}{\mathcal{T}}

\newcommand{\bea}{\begin{eqnarray}}
\newcommand{\eea}{\end{eqnarray}}
\newcommand{\bean}{\begin{eqnarray}}
\newcommand{\eean}{\end{eqnarray}}

\newcommand{\bu}{{\bm u}}

\title[Flow organization in Rayleigh-B\'enard convection]
{Flow organization in non-Oberbeck-Boussinesq Rayleigh-B\'enard convection in water}
\author[K. Sugiyama et al.]{
Kazuyasu Sugiyama$^1$\footnote{Present Address: Department of Mechanical Engineering, School of Engineering,
The University of Tokyo, 7-3-1 Hongo, Bunkyo-ku, Tokyo 113-8656, Japan.}, Enrico Calzavarini$^1$\footnote{Present Address: Laboratoire de Physique de \'Ecole Normale Sup\'erieure de Lyon and  CNRS, 46 All\'ee d'Italie, 69007 Lyon, France.},
Siegfried Grossmann$^2$, and Detlef Lohse$^1$\footnote{d.lohse@utwente.nl}}
\affiliation{
$^1$Physics of Fluids group, Department of Applied Physics, J. M. Burgers Centre for Fluid Dynamics, and Impact-, MESA-, and BMTI-Institutes, University of Twente, P. O. Box 217, 7500 AE Enschede, The Netherlands,\\
$^2$Fachbereich Physik der Philipps-Universitaet, Renthof 6, D-35032
Marburg, Germany
}

\date{\today}
\begin{document}

\maketitle

\begin{abstract}
Non-Oberbeck-Boussinesq (NOB) effects on the flow organization in two-dimensional Rayleigh-B\'enard turbulence 
are numerically analyzed. The working fluid is water. We focus on the 
temperature profiles, the center temperature, 
the Nusselt number, and on the analysis of the velocity field. Several velocity amplitudes (or Reynolds numbers)
and several kinetic profiles 
are introduced and studied; 
these together describe the various features of the rather complex flow organization.   
The results are presented both as functions of the Rayleigh number $Ra$ (with $Ra$ up to $10^8$) 
for fixed temperature difference $\Delta$ between top and bottom plates 
and as functions of $\Delta$ (``non-Oberbeck-Boussinesqness'') for fixed $Ra$
with $\Delta$ up to $60$K. All results are consistent with the available experimental NOB data for 
the center temperature $T_c$ and the Nusselt number ratio $Nu_{NOB}/Nu_{OB}$ (the label OB meaning that the 
Oberbeck-Boussinesq conditions are valid). For the temperature profiles we find -- due to plume emission from 
the boundary layers -- increasing deviations from 
the extended Prandtl-Blasius boundary layer theory presented in (Ahlers et al. 2006 J. Fluid Mech. {\bf 569}, 409--445) 
   with increasing $Ra$, while 
the center temperature itself is surprisingly well predicted by that theory. 
For given non-Oberbeck-Boussinesqness $\Delta$ both the center temperature $T_c$ 
and the Nusselt number ratio $Nu_{NOB}/Nu_{OB}$ only weakly depend on $Ra$. 

Beyond $Ra\approx 10^6$ the flow consists of a large diagonal center convection roll and two smaller rolls in the upper and lower corners, respectively (``corner flows''). In the NOB case the center convection roll is still characterized by only one velocity scale.  In contrast, the top and bottom corner flows are then of different strengths, the bottom one being a factor $1.3$ larger (for $\Delta = 40$K) than the top one, due to the lower viscosity in the hotter bottom boundary layer. Under NOB conditions the enhanced lower corner flow as well as the enhanced center roll lead to an enhancement of the volume averaged energy based Reynolds number $Re^E = \left< \frac{1}{2} \bu^2 \right>^{1/2} L/\nu$ of about $4\%$ to $5\%$ for $\Delta = 60$K. Moreover, we find $Re^E_{NOB}/Re^E_{OB} \approx (\beta (T_c)/\beta (T_m))^{1/2}$, with $\beta$ the thermal expansion coefficient and $T_m$ the arithmetic mean temperature between top and bottom plate temperatures. This corresponds to the ratio of the free fall velocities at the respective temperatures. By artificially switching off the temperature dependence of $\beta$ in the numerics, the NOB modifications of $Re^E$ is less than $1\%$ even at $\Delta = 60$K, revealing the temperature dependence of the thermal expansion coefficient as the main origin of the NOB effects on the global Reynolds number in water.
\end{abstract}
\section{Introduction}\label{sec1}
Rayleigh-B\'enard (RB) convection in the Oberbeck-Boussinesq (OB) approximation (\cite{obe79,bou03})
is rather a mathematical concept than physical reality. On the one hand it is driven by a temperature 
difference $\Delta = T_b - T_t$  between the bottom and top plates, whose distance in height is $L$;  
on the other hand, the temperature dependences of the material properties 
such as the kinematic viscosity $\nu$, the thermal diffusivity $\kappa$, 
the heat conductivity $\Lambda$, the isobaric specific heat capacity $c_p$, 
and the isobaric thermal expansion coefficient $\beta$ are all ignored in the OB approximation 
apart from that of the density $\rho$, which is assumed to vary linearly with $T$,
\be 
\rho(T) \approx  \rho_m (1 - \beta_m (T-T_m)).
\label{rho-lin}
\ee
Here, $T_m = (T_t+T_b)/2$ is the arithmetic mean value of the warmer bottom plate temperature $T_b$ 
and the colder top plate temperature $T_t$, and $\rho_m$ and $\beta_m$ 
denote the density and the thermal expansion coefficient at the mean temperature $T_m$, respectively. 
Fortunately, the OB approximation is rather good, if the material properties depend on temperature only weakly 
or if the temperature difference $\Delta$ between the bottom and top plates is kept small enough. 
However, if either the material properties are strongly temperature dependent (as 
e.g.\ the viscosity of glycerol in some temperature regime) or if the temperature difference between bottom and 
top plates is chosen to be large in order to achieve larger Rayleigh numbers, 
the deviations from the  OB approximation are expected to become relevant. The consequences of these deviations 
are called non-Oberbeck-Boussinesq (NOB) effects, see \cite{ahl06,ahl09}. To which extent they affect 
the Nusselt number and possibly could account for the differences between the Oregon and the Grenoble data 
is an ongoing controversy (\cite{cha97,cha01,roc01,roc02,nie00,nie01,nie03,ash99}).

The signatures of NOB effects studied in this article are (i) a deviation of the center (or bulk) temperature $T_c$ 
from the arithmetic mean temperature $T_m$, (ii) a modified $z$-profile of the area averaged temperature, which develops 
a bottom-top-asymmetry, (iii) different thermal boundary layer thicknesses $\lambda_b \neq \lambda_t$ at the bottom 
and top together with different temperature drops $\Delta_b \neq \Delta_t$ across these BLs, (iv) a modification 
of the Nusselt number, best expressed by the ratio $Nu_{NOB}/Nu_{OB}$, and (v) a bottom-top symmetry broken flow 
structure, in particular possibly different amplitudes $U_b \neq U_t$ of the plate-parallel winds near the bottom and 
top plates meaning different Reynolds numbers $Re_{b,t} = U_{b,t} / \nu L^{-1}$. 

The deviation $T_c - T_m$ of the center temperature from the arithmitic mean temperature and the corresponding
differences between the temperature drops $\Delta_t$ and $\Delta_b$ over the thermal BLs presumably is the 
experimentally most explored NOB effect, namely for water by \cite{wu91a} and by \cite{ahl06,ahl07,ahl08,ahl09}. 
While for water even for $\Delta = 50$~K the deviation $T_c-T_m$ is at most 2 K, for glycerol this deviation can be 
as large as 8 K (\cite{zha97}, \cite{sug07b}). 

To theoretically account for these deviations, in \cite{ahl06} the Prandtl-Blasius BL theory 
was extended to the NOB case, giving surprisingly good agreement with the experimental center temperatures $T_c$ 
for water. Also for ethane gas, which is compressible, the BL theory -- extended to compressible fluid flow -- can
describe the measured center temperature data rather satisfactorily, as presented in \cite{ahl07}. But as was 
shown in \cite{ahl08}, the $T$-dependence of the buoyancy caused by thermal expansion $\beta = \beta(T)$ is 
the dominant cause of the observed NOB effects, in particular the characteristic differences between the more 
gas-like and the more liquid-like ethane on the two sides of the critical isochore.   

The success of the Prandtl-Blasius BL theory  in the context of NOB convection is remarkable for at least three 
reasons: First, the boundary layer theory deals with semi-infinite plates, while experiments are done in finite 
aspect ratio containers, mainly for $\Gamma = 1$. -- ~Second, and more importantly, the Prandtl-Blasius BL theory 
completely ignores the plume separations 
and the corresponding time dependence of the boundary layer flow. Although the shear Reynolds numbers in the BLs 
are not yet very large in RB flow (for $Pr=1$ the transitional shear Reynolds number $Re^S_{\star} \approx 420$, which indicates the range of turbulence transition, is only reached near $Ra\approx 10^{14}$, see \cite{gro02}), the plumes 
(and thus the time dependence of the BL flow) play a significant role in the heat transfer (\cite{cil96,cil99})
and perhaps also for the bulk temperature $T_c$. Plumes are not included in the classical BL theory because that does not take notice of the buoyant forcing in the Prandtl approximation of the hydrodynamic equations of motion. Neither the thermal expansion coefficient $\beta$ itself is addressed nor is 
its temperature dependence taken into account, which in reality is considerable. For water at $T_m=40^o$C 
and $\Delta = 40$K there is nearly a factor of 2 between the respective values $\beta_t$ and $\beta_b$ at the top and bottom plates. 
Technically speaking,  the BL theory misses $\beta$ since buoyancy shows up in the equation of the 
vertical velocity field $u_z$; and this equation enters into the Prandtl approximation only to derive the height 
independence of the pressure, which in RB anyhow does not play a role. The numerical simulations presented in 
this paper will quantitatively show that buoyancy and plumes indeed affect the temperature BL profiles. 

\begin{figure}
\begin{center}
\vspace*{0.3cm}
\epsfig{file=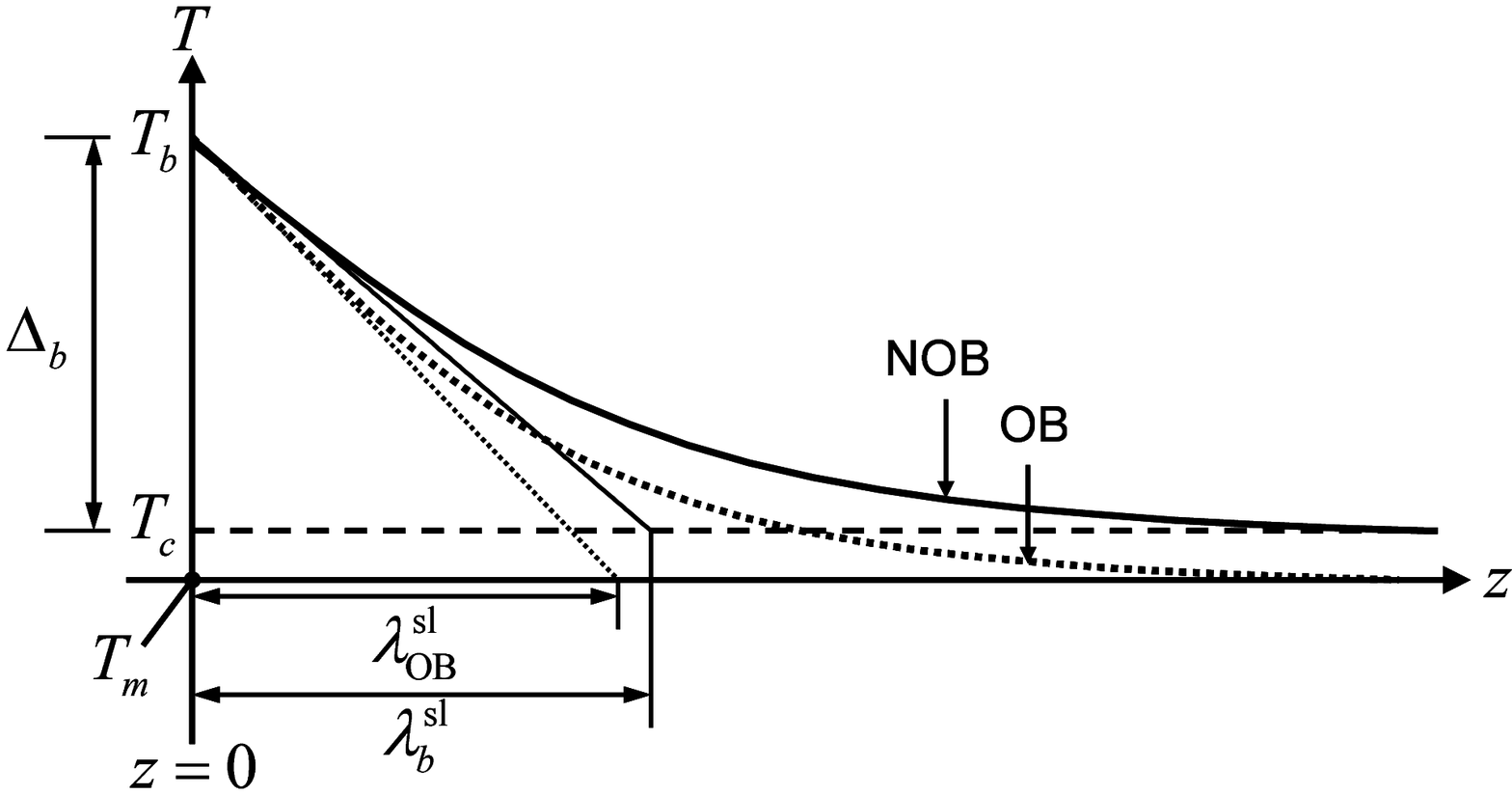,width=.75\textwidth}
\hspace*{0.3cm}\epsfig{file=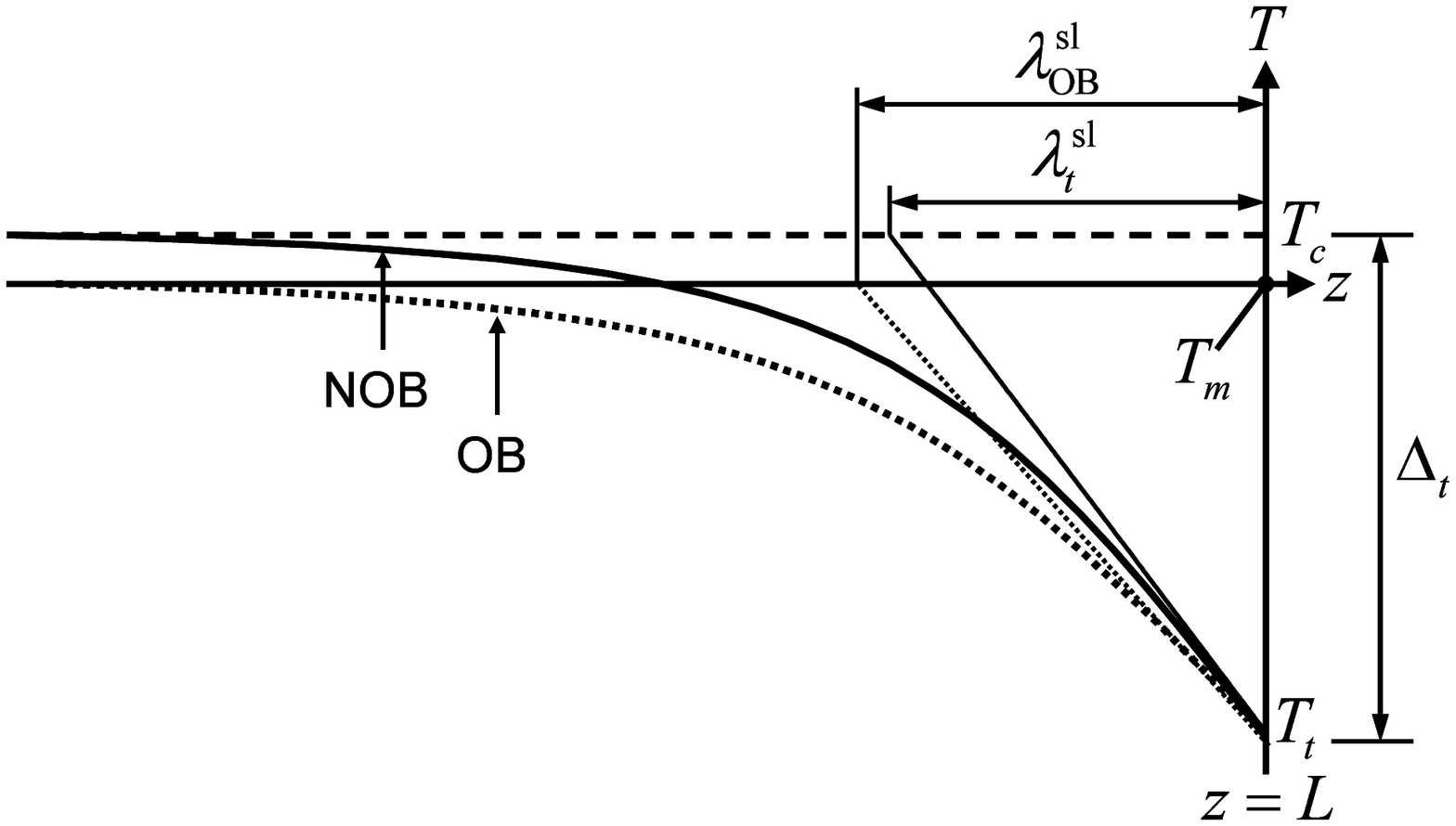,width=.75\textwidth}
\vspace*{0.3cm}
\end{center}
\caption{
Notations and sketch of the time and surface averaged temperature profiles versus height $z$ in the OB and 
NOB cases, respectively. The height of the cell is $L$ (not shown). The temperatures at the top $(z=L)$ and the 
bottom $(z=0)$ plates are $T_t$ and $T_b$, their arithmetic mean is $T_m=(T_t+T_b)/2$.
The thermal BL thicknesses based on the temperature slopes at the plates are called $\lambda_{t,b}^{sl}$.  
The respective temperature drops are $\Delta_t$ and $\Delta_b$. The time averaged temperature in the bulk (or 
center) is $T_c$. For water as the working fluid this center temperature $T_c$ is larger than  
$T_m$. While $\lambda_b^{sl} = \lambda_t^{sl}$ in the OB case, under NOB conditions 
the bottom BL is thinner than the top one, $\lambda_b^{sl} < \lambda_t^{sl}$. The fluid properties such as $\nu$, $\kappa$, 
and $\beta$ carry the same index as the temperature at which they are considered, e.g. 
$\nu_t = \nu (T_t)$ for the  kinematic viscosity at the top plate, and so on.
} 
\label{sketch_bl}
\end{figure}

There are two further assumptions of the extended BL theory developed in \cite{ahl06} which
need to be tested: First, the extended BL theory assumes that the large scale wind velocity is the same close to
the top and the bottom plate, i.e., $U_t = U_b$ or $Re_t = Re_b \equiv Re_{NOB}$. We will show that there are several 
relevant velocity amplitudes, most of which break the top-bottom symmetry. Only the main central roll has the same 
amplitude near the bottom as well as the top boundary layers. Second, within the BL theory the ratios 
$Nu_{NOB}/Nu_{OB}$ and $Re_{NOB}/Re_{OB}$ cannot be calculated at all, in contrast to $T_c$. These quantities can  
only be obtained by additional input from experiment, namely by employing the experimental information on the ratio 
$F_\lambda := {2\lambda_{OB}^{sl} / ( \lambda_t^{sl} + \lambda_b^{sl})}$. Here $\lambda_{t}^{sl}$ and 
$\lambda_{b}^{sl}$ are the top and bottom thermal BL thicknesses, defined via the temperature slopes at the plates
in Eq. (\ref{thicknesses}) and sketched in figure \ref{sketch_bl}. In \cite{ahl06} we have calculated 
$F_\lambda$ from the measured Nusselt number and the calculated center temperature. Its value $T_c$ determines the 
ratio $F_\Delta := {(\kappa_t \Delta_t + \kappa_b \Delta_b) / (\kappa_m \Delta )}$). The exact relation holds (even 
for compressible flow){\footnote{The notation used in eq.\ (\ref{nu_ratio}) is explained in the caption of figure
\ref{sketch_bl}.}}
\be
\frac{Nu_{NOB}}{Nu_{OB}}
= \frac{ 2 \lambda_{OB}^{sl}}{ \lambda_t^{sl} + \lambda_b^{sl} } \cdot 
\frac{ \kappa_t \Delta_t + \kappa_b \Delta_b }{ \kappa_m \Delta }
=: F_{\lambda} \cdot F_{\Delta}.
\label{nu_ratio}
\ee
$Nu_{NOB}$ is the actual heat flux with all material parameters taken at their respective real temperature values. The 
label OB means that all fluid properties are taken as temperature independent constants, evaluated 
at the arithmetic mean temperature $T_m$. 

Remarkably, for the analyzed case of water at $T_m=40^o$C (see \cite{ahl06}) the experimental data are consistent 
with $F_\lambda = 1$. But as was shown in \cite{sug07b} by numerical simulations it is $F_{\lambda} \neq 1$ for 
RB convection in glycerol under NOB conditions, at least up to $Ra=10^8$. However, in glycerol due to the large 
Prandtl number $Pr = 2500$ a large scale 
convection roll did not yet develop and it could be that $F_\lambda = 1$ is connected with the existence of such a 
roll. The numerical simulations presented in this paper will unambiguously show that $F_\lambda = 1$ 
does {\it not} hold in general. This property thus is co-incidental for water due to the specific temperature 
dependences of its material properties around $40^o$C. 

The question of modifications of the Reynolds number(s) through NOB effects is intimately related to the flow 
organization. In recent years there was considerable progress in our 
insight into the flow structure, thanks to numerical simulations (see e.g.\ \cite{ver03b,sch04,ama05,str06}),
to PIV measurements  (\cite{xia03,qiu04,xi04,sun05,sun05a}), and to velocity correlation measurements 
(\cite{bro07b}). These papers revealed that there are various feasable possibilities to define flow amplitudes 
and that these differently defined amplitudes and the corresponding Reynolds numbers have different scaling behavior 
with $Ra$. Our numerical simulations have fully confirmed and detailed this view. We will show that NOB conditions 
influence the flow structure near top and bottom differently and modify the various Reynolds numbers correspondingly. 
NOB conditions have 
the largest impact on the convective flow in one top and one bottom corner of the cell, where macroscopically 
visible secondary rolls develop. We will also show that the NOB modification of the global, volume and time averaged, 
energy based Reynolds number $Re^E = \left< \frac{1}{2}\bu^2\right>_{V,\tau}^{1/2} /(\nu L^{-1})$ is consistent with 
attributing it mainly to the change of the thermal expansion coefficient $\beta$ in the bulk. More specifically, 
we find $Re^E_{NOB}/Re^E_{OB} \approx (\beta (T_c)/\beta (T_m) )^{1/2}$, a finding clearly not describable within 
the extended Prandtl-Blasius BL theory.

In this paper we focus on water ($Pr=4.4$). Nevertheless, the parameter space is considerable.
Next to $Ra$ the crucial parameter is the NOBness $\Delta$. For comparison we perform numerical simulations for
fluids with non-physical temperature dependences of their material properties in order to clarify the origin of 
certain observations. 

As the numerical effort is so large for three-dimensional simulations we restrict ourselves to two-dimensional 
simulations. One may worry on whether two-dimensional simulations are sufficient to reflect the dynamics of  
three-dimensional RB convection. For heat flow under OB conditions this point has been analyzed in detail by 
\cite{sch04} and earlier by \cite{delu90,wer91,wer93}. \cite{sch04}'s conclusion is that for $Pr\ge 1$ 
various properties observed in numerical 3D convection (and thus also in experiment) are well reflected in the 2D 
simulations. This in particular holds for the BL profiles and for the Nusselt numbers. Since one focus of this paper 
is on the difference between OB and NOB convection, the restriction to 2D simulations might be even less severe, 
as NOB deviations occur in both cases and the differences between 2D and 3D simulations might  
cancel out in quantities such as $(T_c - T_m )/ T_m$, $Nu_{NOB}/Nu_{OB}$, or $Re_{NOB}/Re_{OB}$. We also note that 
for a comparison with the Prandtl-Blasius BL theory 2D simulations are in fact more appropriate than 3D 
simulations, as the BL theory is two-dimensional per construction. 

The paper is organized as follows: 
In section \ref{sec2} we will explain, justify, and verify the numerical method. Section \ref{sec3} is devoted to 
our results on the mean temperature profiles and the related shifts of the center temperatures. Section \ref{sec4} 
addresses the NOB effects on the Nusselt number. The main  section is section \ref{sec5} where we first analyze the 
flow structure for the OB case and then its modifications through NOB effects. Several feasable measures for the 
wind amplitudes of the complex flow structure will be introduced. Section \ref{sec6} contains the conclusions.

\section{Definitions, governing equations, and numerical method} \label{sec2}

The equations governing non-Oberbeck-Boussinesq convection in incompressible liquids are the 
incompressibility condition
\be
\partial_i u_i=0 \ ,
\label{incompressibility}
\ee 
the Navier-Stokes equation
\be
\rho_m (\partial_t u_i + u_j \partial_j u_i) = 
-\partial_i p  + \partial_j (\eta ( \partial_j u_i +\partial_i u_j )) + \rho_m g 
\left( 1 - \frac{\rho}{ \rho_m} \right)  \delta_{i3} ,
\label{ns}
\ee
and the heat-transfer equation
\be
\rho_m c_{p,m} 
(\partial_t T + u_j \partial_j T ) = 
\partial_j ( \Lambda \partial_j  T ) .
\label{ht}
\ee
The dynamic viscosity $\eta(T)$ and the heat conductivity $\Lambda(T)$ are both temperature and thus space dependent. 
The isobaric specific heat capacity $c_p$ and the density $\rho$ are assumed as constants and their values 
$c_{p,m}, \rho_m$ fixed at the temperature $T_m$, except in the buoyancy term, where the full nonlinear 
temperature dependence of $\rho(T)$ is implemented. For water, which we here consider as the working fluid, 
density as well as specific heat are indeed constant with temperature to a very good approximation. The 
experimentally known temperature dependences of $\eta$, $\Lambda$, and $\rho$ (in the buoyancy term) together with
the values of the parameters  
$\rho_m$, $c_{p,m}$ for water are given in the appendix of \cite{ahl06} and, for better reference, are reported in 
Tab. \ref{tab1} in the form implemented in the present DNS.

\begin{table}

\begin{center}

\begin{tabular}{lrlrlrl}
&$\nu$&&$\kappa$&&$g(1-\rho/\rho_m)$&\\
$n$&$A_n$&&$B_n$&&$C_n$&\\\hline
0&$ 6.6945204\cdot 10^{-7 }$&[m$^2$/s]         &$ 1.5222630\cdot 10^{-7 }$&[m$^2$/s]	     & 0                   &[m$^2$/s]                        	          \\
1&$-1.215394\cdot 10^{-8 } $&[m$^2$/(s\ K)]    &$ 3.347639 \cdot 10^{-10}$&[m$^2$/(s\ K)]    &$ 3.7576156\cdot 10^{-3}$&[m/(s$^2$\ K)]    \\
2&$ 1.737730\cdot 10^{-10} $&[m$^2$/(s\ K$^2$)]&$-2.702875 \cdot 10^{-12}$&[m$^2$/(s\ K$^2$)]&$ 3.900878 \cdot 10^{-5}$&[m/(s$^2$\ K$^2$)]\\
3&$-2.48455\cdot 10^{-12}  $&[m$^2$/(s\ K$^3$)]&$0$		        &		     &$-1.811623 \cdot 10^{-7}$&[m/(s$^2$\ K$^3$)]\\
4&$ 3.55232\cdot 10^{-14}  $&[m$^2$/(s\ K$^4$)]&$0$		        &	  	     &$0$		       &                  \\
5&$-5.0790\cdot 10^{-16}   $&[m$^2$/(s\ K$^5$)]&$0$                     &                    &$0$                      &                  \\
\hline
\end{tabular}

\end{center}

\caption{ 
Expansion coefficients of material properties of water around the temperature 
$T_m = 40^{\rm o}$C adapted from \cite{ahl06}. 
The kinematic viscosity, the thermal diffusivity, and buoyancy are written in a polynomial form as 
$\nu(T) \equiv \eta(T)/\rho_m$$=$$\sum_{n=0}A_n (T-T_m)^n$, 
$\kappa(T) \equiv \Lambda(T)/(\rho_m c_{p,m})$ $=$ $\sum_{n=0}B_n (T-T_m)^n$, 
and $g(1-\rho(T)/\rho_m)$$=$$\sum_{n=0}C_n (T-T_m)^n$, respectively.
Using the leading coefficient $C_1 (=\beta_m)$ for the buoyancy force, we can write the Rayleigh number 
defined in (\ref{ra-def}) as $Ra=C_1 
L^3\Delta/(\nu_m \kappa_m)$, where $\nu_m=\eta(T_m)/\rho_m$ and $\kappa_m=\Lambda(T_m)/(\rho_m c_{pm})$, 
which coincides with the usual OB definition.}
\label{tab1}
\end{table} 
 
We deal with a wall-bounded system with an aspect ratio fixed at $\Gamma = 1$.
The velocity boundary conditions accompanying the dynamical equations are $u_i=0$ at the top and bottom 
plates $z=L$ and $z=0$ as well as on the side walls $x=0, x=L$. The temperature boundary conditions are 
$T_b-T_t = \Delta$ for the temperature drop across the whole cell of height $L$. At the side-walls 
($x=0, ~x=L$) heat-insulating conditions are employed, $\partial_x T |_{x=0,L} =0$. 
The cell is considered to be 2-dimensional, i.e., there is no $y$-dependence.
The Rayleigh number is defined with the material parameters taken at the mean temperature $T_m$,
\be
Ra = \frac{ \beta_m g L^3 \Delta }{\nu_m \kappa_m }.
\label{ra-def}
\ee
We vary the Rayleigh number in DNS by varying the height $L$ of the box, 
while the non-Oberbeck-Boussinesq\-ness is changed by varying the temperature drop $\Delta$. 
Note that in eq.\ (\ref{ns}) the full temperature dependence 
of the density in the buoyancy term is taken into account,
rather than employing the linear approximation eq.\ (\ref{rho-lin}) only. 
Still the Rayleigh number is defined with the coefficient of the linear expansion 
of the density with respect to temperature, taken at the mean temperature, $\beta_m = \rho_m^{-1} 
\partial \rho / \partial T|_{T_m}$, cf. Tab. \ref{tab1}. The Prandtl number $Pr = \nu_m /\kappa_m$ is 
also defined in terms of the 
material parameters at the arithmetic mean temperature. 

Equations (\ref{incompressibility})-(\ref{ht}) are solved on a two-dimensional domain with gravity pointing in 
negative $z$-direction. To discretize the Navier-Stokes and heat transfer equations, we employ a finite difference 
scheme (see e.g. \cite{pey83}; \cite{fer96}). The space derivatives are approximated 
by the fourth-order central difference scheme on a staggered grid (\cite{har65}). In particular for the advection 
terms we employ the scheme proposed by \cite{kaj01}, which satisfies the relations 
$\partial_j(u_ju_i)=u_i\partial_ju_j+u_j\partial_j u_i$ and $\partial_j(u_jT)=T\partial_ju_j+u_j\partial_j T$
in a discretized form and ensures that the second moments of the velocity and temperature are highly conserved. 
To integrate the equations in time, we use the second-order scheme, i.e., the Adams-Bashforth method for the 
advection terms and the Crank-Nicolson one for the viscous, diffusive, and buoyant terms (see e.g. \cite{can88}).  
To complete the time marching in the momentum equation and simultaneously satisfying the solenoidal condition 
(\ref{incompressibility}) of the velocity vector, we employ a simplified-marker-and-cell procedure (\cite{ams70})
by solving a Poisson equation for the pressure. The two-dimensional discretized pressure equation, 
which is written in the fourth-order finite difference form, is reduced into a one-dimensional problem by taking the 
Fast Fourier Transform (FFT) in the $x$-direction. The boundary condition at the side walls ($x=0$ and $x=L$) is 
satisfied, if the relation $\partial_x\phi|_{x=0}=\partial_x\phi|_{x=L}=0$ holds for all the quantities $\phi$ 
in the pressure equation. To impose this condition, we take a periodicity $2L$ for the FFT 
(i.e., $\phi(x)=\phi(2nL+x)$ with $n$ an arbitrary integer) and introduce a fictitious domain $L<x\leq 2L$, 
in which the quantities are given by $\phi(x)=\phi(2L-x)$. We directly solve the reduced-order equation 
written in a heptadiagonal matrix form and then determine the pressure field by taking the inverse FFT. 

We have validated the numerical code by checking the instantaneous kinetic energy and entropy budget relations for 
$\bu^2 /2$ and $T^2/2$ both in the OB and NOB cases and by evaluating the correctness of the onset of convection 
in the OB case. The critical Rayleigh number we compute ($Ra_c=2585.27$) is in agreement with the one computed 
analytically by \cite{luij81} ($Ra_c=2585.03$) to a precision of less than $0.01\%$. We note that $Ra_c$ is much 
larger than the more known $Ra_c=1708$ for an infinite aspect ratio system, due to the presence of lateral walls. 

\begin{figure}
\begin{center}
\vspace*{0.3cm}
\epsfig{file=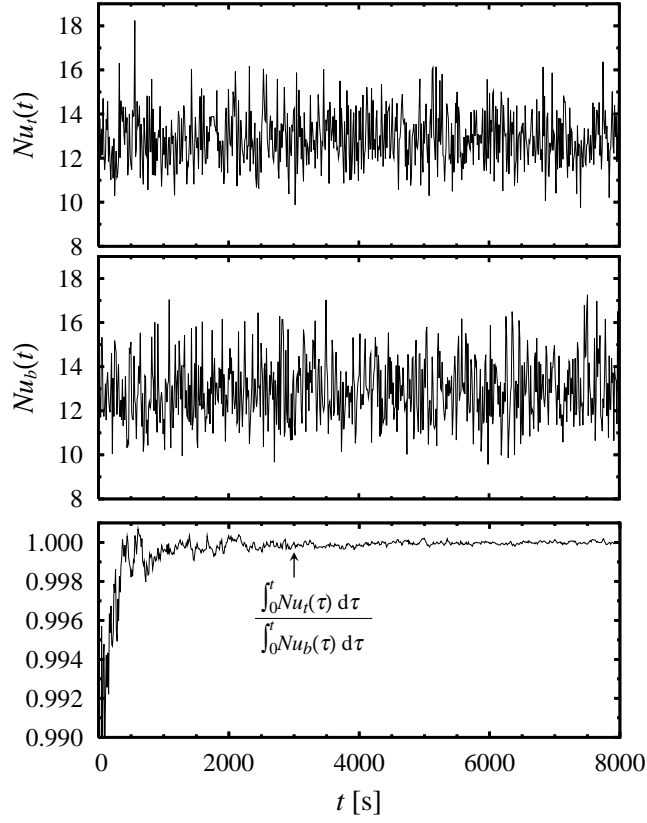,width=11cm,angle=-90}
\vspace*{0.3cm}
\end{center}
\caption{
Upper two panels: 
Temporal evolution of $Nu_t(\tau)$ and $Nu_b(\tau)$ as defined in equation (\ref{nu-top-bot}), before time averaging,
for $Ra=10^7$, $\Delta=40$K, and $L=1.89\ $cm. The total integration time $\bigT$, reported here in seconds, is of 
the order $10^4$ integral times of the autocorrelation coefficient, $\tau_I \simeq  0.606\ $s. 
Lowest panel:
Ratio of the temporally averaged Nusselt numbers $\int_{0}^{t}Nu_t (\tau) \ {\rm d}\tau /\int_{0}^{t}Nu_b(\tau)\ 
{\rm d}\tau$. Here the time averages at top and bottom agree up to $0.009$\%. For other $Ra$ and $\Delta$ the 
statistical convergence is similar with maximum relative error being $0.091$\%.
} 
\label{nu-t+nu-b}
\end{figure}

The area averaged heat currents are calculated as functions of time $\tau$ both at the top ($t$) and the 
bottom ($b$) plates separately,
\begin{equation}
Nu_t(\tau) = \frac{ -\kappa_t ~\partial_z \left<T \right>_{A_t} (z=L, \tau )}{\kappa_m \Delta / L}~, \qquad 
Nu_b(\tau) = \frac{-\kappa_b ~\partial_z \left<T \right>_{A_b} (z=0, \tau )}{\kappa_m\Delta/L}~,
\label{nu-top-bot}
\end{equation}
with $\langle \ldots \rangle_A$ denoting the averaging over the horizontal surface (actually over $x$ only,
since the system is 2D) of the top and bottom plates. We find good agreement of the time averages of 
$Nu_t(\tau), Nu_b(\tau)$, see figure \ref{nu-t+nu-b}. 

To quantify the statistical convergence, we make an uncertainty analysis estimating the time autocorrelations (see 
for instance \cite{ten72} Sect. 6.4). For a time-dependent function $f(\tau)$, the error is evaluated as 
$\delta f=f_{\rm rms}\sqrt{2\tau_I/\bigT}$; here $f_{\rm rms}$ is the root mean square of $f$, $\tau_I$ the 
integral time obtained from the autocorrelation coefficient of $f$, and $\bigT$ the total simulation time under 
statistically steady conditions, i.e., after a transient time $\tau_0$. Considering the error propagation, we 
evaluate the errors of $F_{\lambda} \cdot F_{\Delta}$, of $F_{\lambda}$, and of $F_{\Delta}$ at 95\% 
confidence level as, respectively,
\begin{equation}\begin{split}
\delta (F_{\lambda}F_{\Delta})
=&\frac{Nu_{NOB}}{Nu_{OB}}
\sqrt{\left(\frac{\delta(Nu_{NOB})}{Nu_{NOB}}\right)^2+
\left(\frac{\delta(Nu_{OB})}{Nu_{OB}}\right)^2},\\
\delta (F_{\lambda})
=&F_{\lambda}
\sqrt{\left(\frac{\delta(Nu_{NOB})}{Nu_{NOB}}\right)^2+
\left(\frac{\delta(Nu_{OB})}{Nu_{OB}}\right)^2+
\left(\frac{\delta(F_\Delta)}{F_\Delta}\right)^2},\\
\delta (F_{\Delta})
=&\left|\frac{\kappa_b-\kappa_t}{\kappa_m\Delta}\right|
\delta(T_c),
\end{split}\end{equation}
which will be indicated by the error bars in the plots.

\begin{figure}
\begin{center}
\vspace*{0.3cm}
\epsfig{file=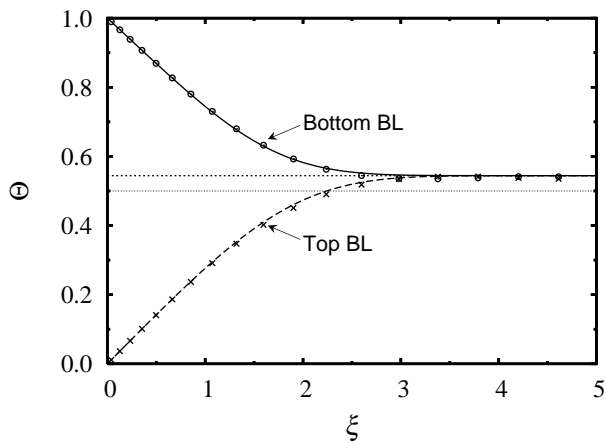,width=6cm,angle=-90}
\vspace*{0.3cm}
\end{center}
\caption{
Temperature profiles $\Theta(\xi) = (\langle T\rangle_{A(\xi)} -T_t)/\Delta$ for $Ra=10^4$ and $\Delta = 40$K 
for water at $T_m=40^o$C. The symbols indicate the results of the numerical simulation, 
the lines stem from the extended Prandtl-Blasius boundary layer theory proposed in \cite{ahl06}. 
The numerical simulations give $\Theta (z)$. $\xi = z/l_{\rm BL}$ is the similarity variable. In classical 
BL theory its scale is given by $l_{\rm BL} = \sqrt{x \nu_m / U_{t,b}}$ with $x$ being the distance from the 
plate's edge; in this figure in order to translate to $\Theta (\xi )$ we have chosen the factor $l_{\rm BL}$ such that the curves have the same slope at $\xi=0$ .
} 
\label{profile-4-40}
\end{figure}

\begin{figure}

\vspace*{0.3cm}
\ \hspace{5em}\ (a)
\vspace*{-2em}
\begin{center}
\epsfig{file=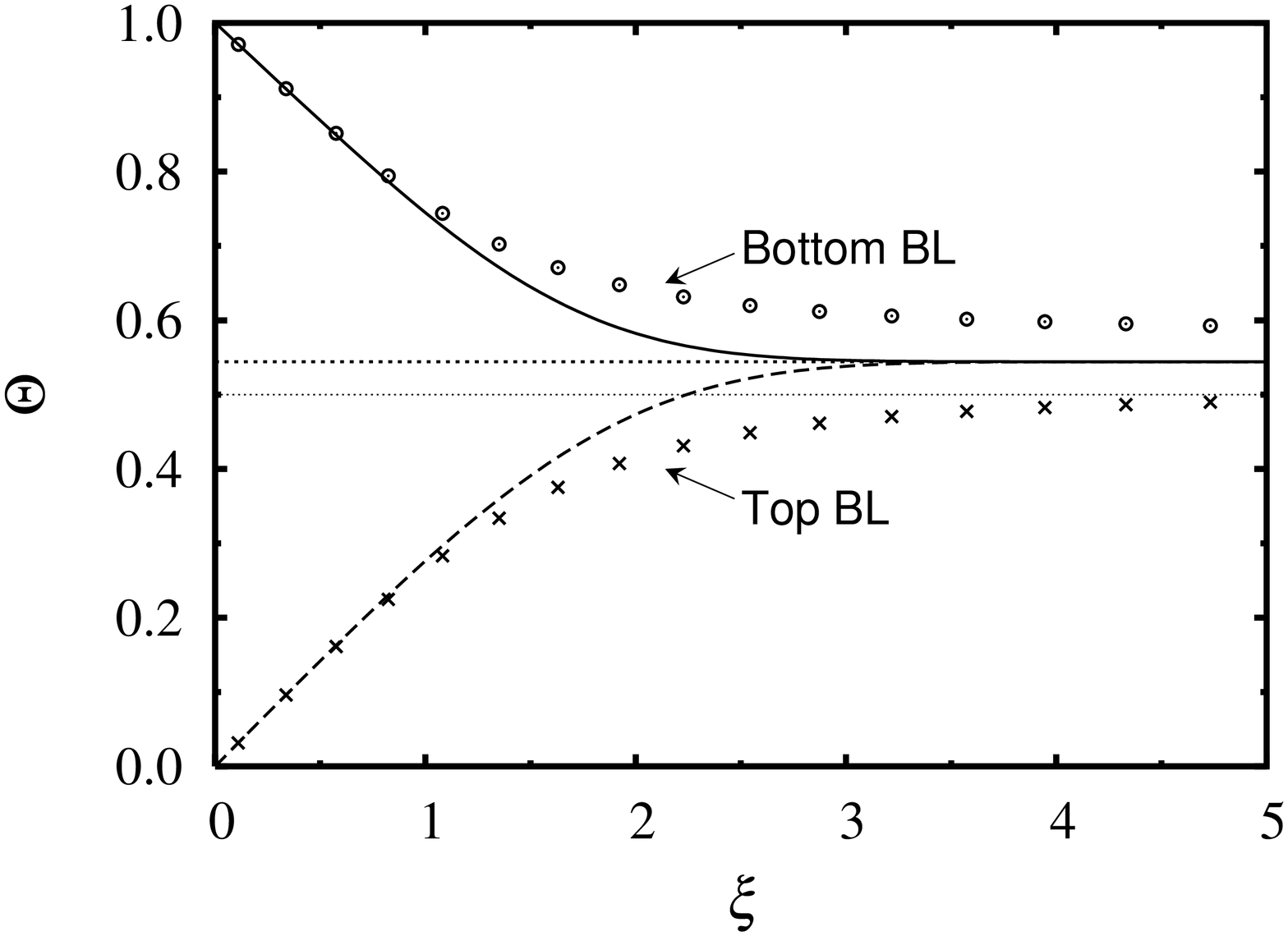,width=6cm,angle=-90}
\end{center}

\vspace*{0.3cm}

\ \hspace{5em}\ (b)
\vspace*{-2em}
\begin{center}
\epsfig{file=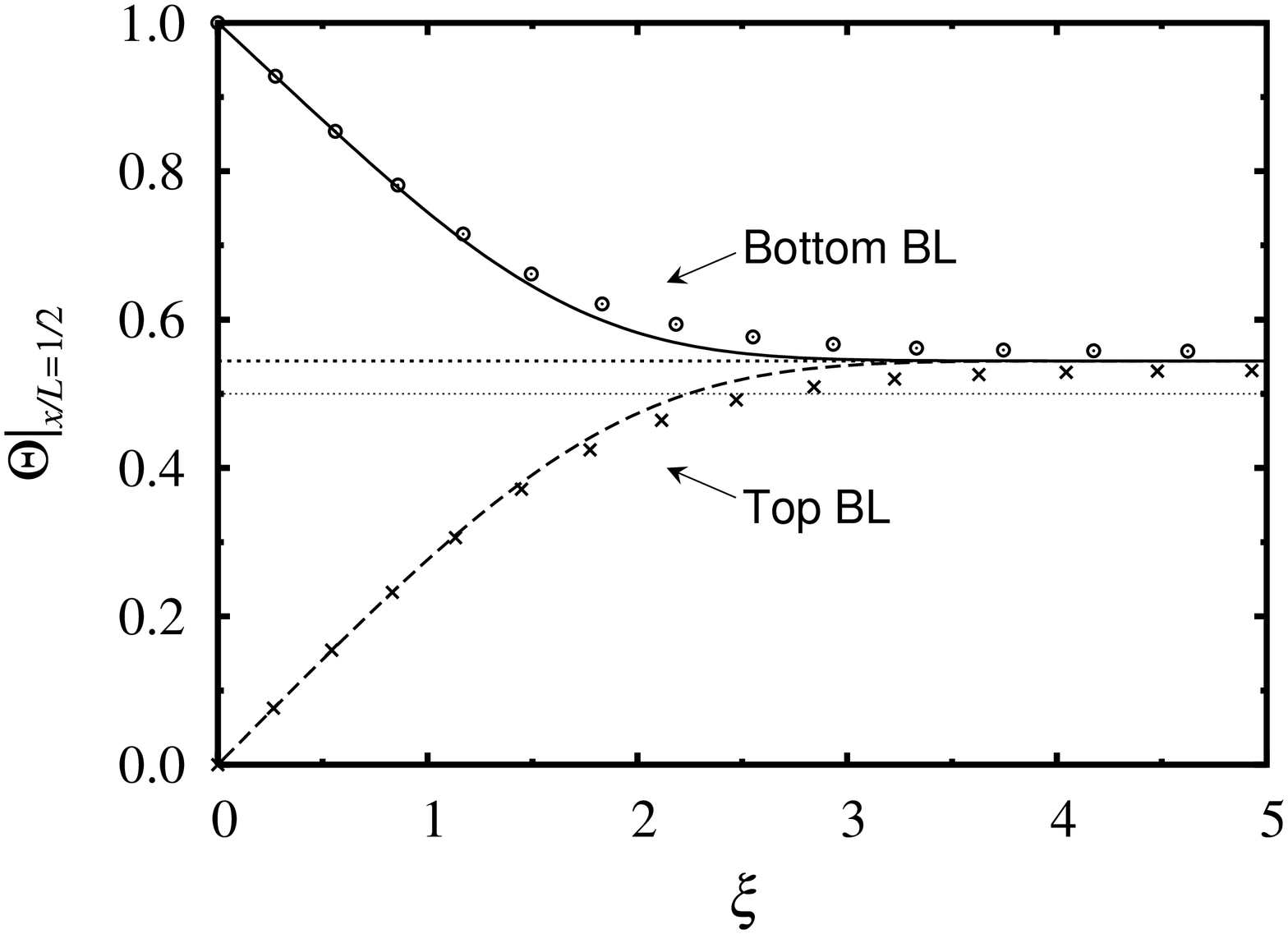,width=6cm,angle=-90}
\vspace*{0.3cm}
\end{center}

\caption{
Temperature profiles at $Ra=10^8$ with $\Delta = 40$K and $T_m=40^o$C in water. 
Upper panel (a): horizontally area averaged temperature profiles as in figure \ref{profile-4-40}.
Though the center temperature $T_c$ is well described by the extended Prandtl-Blasius BL theory, for 
this relatively large $Ra$ the numerical temperature $z$-profiles show significant deviations from those 
of the extended BL theory. 
We attribute these deviations to the plume detachments, which are not included  
in the extended Prandtl-Blasius BL theory. To demonstrate this we show in the lower panel (b): time averaged 
temperature $z$-profiles at a fixed $x$-value, namely along the middle line $x=L/2$, i.e., 
 $\Theta(\xi)|_{x=L/2}=(T|_{x=L/2}-T_t)/\Delta$. Here the plume activity is expected to be  
weaker than in the regions near the side walls, which contribute to the area averaged profile in panel (a). Indeed 
DNS and BL theory agree satisfactorily along the center line. 
} 
\label{profile-8-40}
\end{figure}

\begin{figure}
\begin{center}
\vspace*{0.3cm}
\epsfig{file=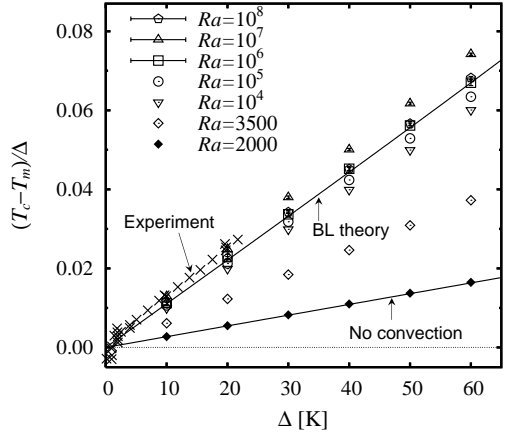,width=6cm,angle=-90}
\vspace*{0.3cm}
\end{center}
\caption{
Relative deviation $(T_c-T_m)/\Delta$ of the horizontally area (and time) averaged center temperature $T_c$ from the 
arithmetic mean temperature $T_m$ in terms of $\Delta$ versus the temperature difference $\Delta$ for water 
at fixed $T_m=40^o$C for various values of $Ra$. The nearly linear increase means that 
$T_c - T_m = \mbox{const} \Delta^2 + h.o.t.$
} 
\label{tc-vs-del}
\end{figure}

\begin{figure}
\begin{center}
\vspace*{0.3cm}
\epsfig{file=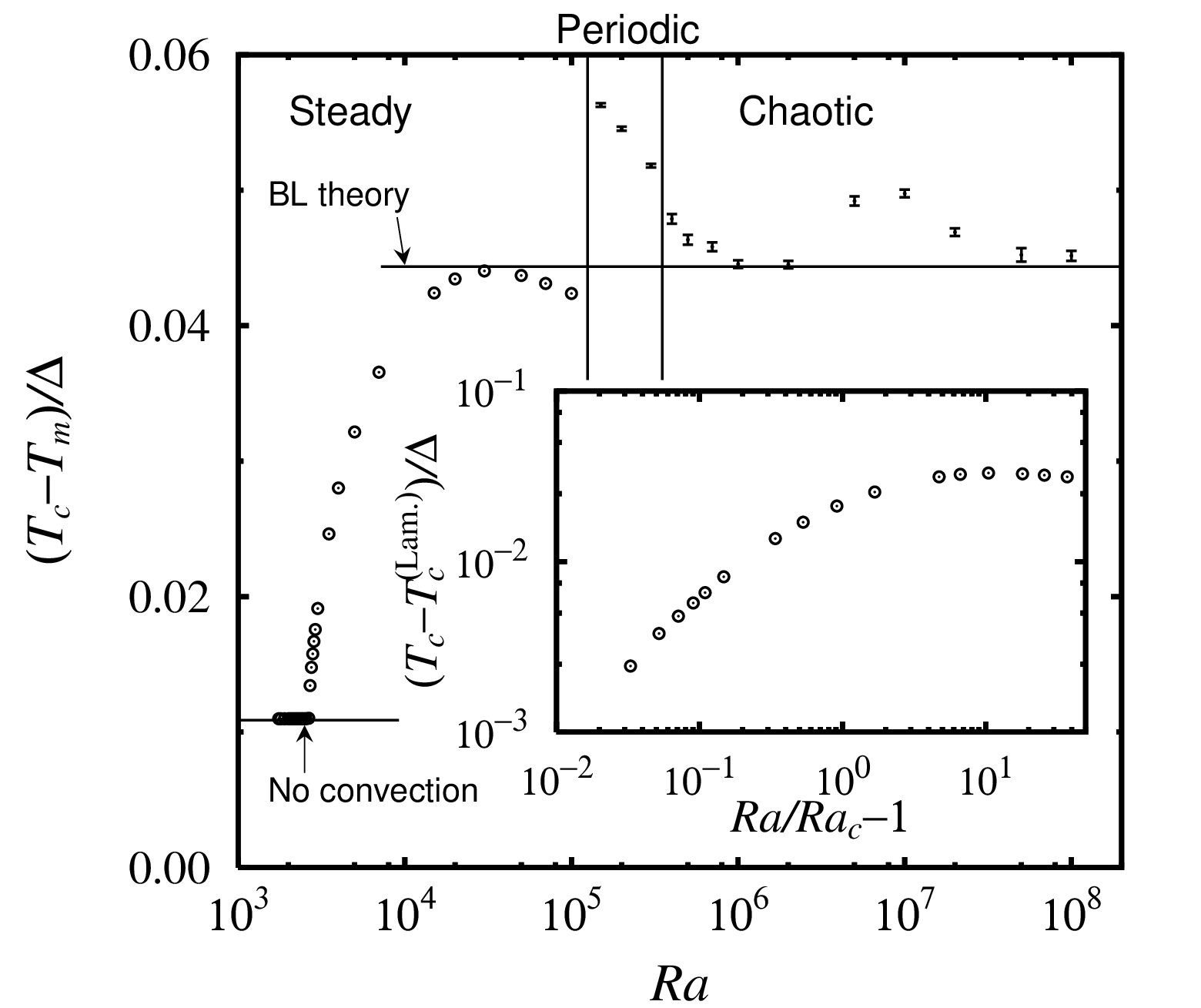,width=6cm,angle=-90}
\vspace*{0.3cm}
\end{center}
\caption{
$(T_c-T_m)/\Delta$ vs $Ra$ for water at fixed $T_m=40^o$C and $\Delta = 40$K. Inset: deviation of $T_c$ from the 
center temperature in the laminar case $T_c^{(lam)}$ at the onset $Ra_c=2590$ of thermal convection at $\Gamma = 1$.
Note the strong variation of $T_c$ in a range slightly above $Ra = 10^5$, which we attribute to transitions between 
different coherent flow structures.
} 
\label{tc-vs-ra}
\end{figure}

\section{Mean temperature profiles and center temperature}\label{sec3}
We will first focus on the water case with $T_m=40^o$C, corresponding to the experiments of \cite{ahl06}. 
In figures \ref{profile-4-40} and \ref{profile-8-40} the mean temperature profiles, averaged over the full width 
$0 \le x \le L$ of the cell, are shown for $Ra=10^4$ and $Ra=10^8$, respectively and compared with the extended 
Prandtl-Blasius BL theory developed in \cite{ahl06} (the specific procedure adopted for the comparison is 
detailed in the caption of Fig. \ref{profile-4-40}).

For $Ra=10^4$ the agreement between numerical data and the time independent extended BL theory is 
excellent, both for the profiles and for the center temperature. This good agreement is remarkable as 
originally the Prandtl-Blasius BL theory has been derived for semi-infinite or at least long flat plates.

For $Ra=10^8$ there are differences between the numerical and the BL theory profiles, namely, the numerical profiles 
are somewhat smoother than those from the BL theory, upper panel (a) of figure \ref{profile-8-40}. 
At this relatively large $Ra$ such differences are not unexpected, because of the enhanced plume activity at larger
$Ra$, which is not included in the extended BL theory. The plume activity is not homogeneous in 
the horizontal direction and the area averages in evaluating $\Theta (\xi)$ in the upper figure
is taken over the whole container width $0 \le x \le L$, including also the neighborhoods of the side walls, where 
the plume convection is preferentially strong. The influence of the plume convection can be confirmed 
by comparing with the numerical, only time averaged $z$-profiles along the middle line $x=L/2$, see lower panel 
of figure \ref{profile-8-40}. These profiles now show good agreement with the extended BL theory , much better
than those in the upper panel. We attribute this to the expectation that the plume detachment near $x=L/2$ is 
less than that near the side walls, and thus the BL approximation should be more reasonable along the middle line.
It could be objected that the observed deviations might be due to the influence of the side-walls, 
which are included in the horizontal surface averages but excluded for the center-line, with only time averaging.
To clarify this point we have performed a DNS with lateral periodic boundary conditions and $\Gamma=2$.
The numerically obtained profiles (not shown here) are very close to the surface and time averaged profiles of 
Fig. \ref{profile-8-40}(a) rather than to the time averaged center line profiles in  Fig. \ref{profile-8-40}(b). 
This supports the conclusion that it is the plume flow and temporal dynamics of the BLs, which is the 
main reason for the observed discrepancy between the Prandtl-Blasius BL theory and the area averaged DNS profiles.

Nevertheless, in spite of the large deviations in the $z$-(or $\xi$-)dependence of the area averaged 
temperature profiles as shown in 
figure \ref{profile-8-40}(a), the center temperature $T_c$ obtained from the extended BL theory (\cite{ahl06}) still 
very nicely agrees with that calculated with DNS. This is confirmed in figure \ref{tc-vs-del}, showing $T_c$ as a 
function of the NOBness $\Delta$ for various $Ra$, ranging from $2\cdot 10^3$ to $10^8$. In figure \ref
{tc-vs-ra} we display the $T_c$-shift for fixed $\Delta = 40$K and $T_m = 40^o$C as a function of $Ra$. 
Interestingly enough, beyond some $10^5$ for $Ra$, the center temperature 
$T_c$ is rather independent of $Ra$. Only in the immediate range beyond the onset of convection, $T_c-T_m$ is 
pronouncedly smaller, reflecting the smooth transition to the small value of $T_c-T_m$ in the non-convecting state. 
The $Ra$ dependence of $T_c$ is not monotonous, what we attribute to transitions between the various coherent 
RB flow patterns in the considered $Ra$-range. 

The results shown in Fig. \ref{tc-vs-ra} are consistent with previous findings (\cite{loh93}) that in the $Ra$-range 
from onset of convection up to about $Ra \approx 5 \cdot 10^7$ the flow only successively looses its spatial coherence. 
In this $Ra$-range the relative coherence length $\ell_{coherence} / L$ decreases from values far above 1 to values 
of order 1/6, see figure 1 of \cite{sug07b} in which we have calculated $\ell_{coherence} / L$ as a function of 
$Ra$, based on the unifying theory of \cite{gro00,gro01,gro02,gro04}. Only for $Ra$ beyond this transition range up to  
some $10^7$, in which spatially coherent structures are gradually lost, the heat convection is fully turbulent and the 
pdf of the fluctuations becomes exponential instead of being Gaussian.

\begin{figure}
\begin{center}
\vspace*{0.3cm}
\epsfig{file=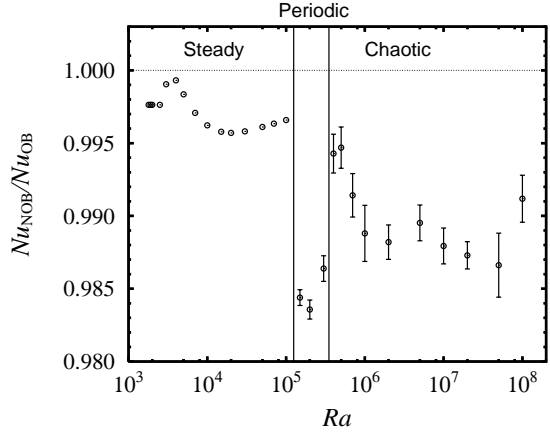,width=6cm,angle=-90}
\vspace*{0.3cm}
\end{center}
\caption{
Nusselt number ratio  $Nu_{NOB}/Nu_{OB}$ vs. Rayleigh number $Ra$ 
for water at fixed values for $T_m=40^o$C and $\Delta = 40$K. 
It is also indicated where we are in the steady, periodic, or chaotic regime.
} 
\label{nu-ratio-vs-ra}
\end{figure}

\begin{figure}
\begin{center}
\vspace*{0.3cm}
\epsfig{file=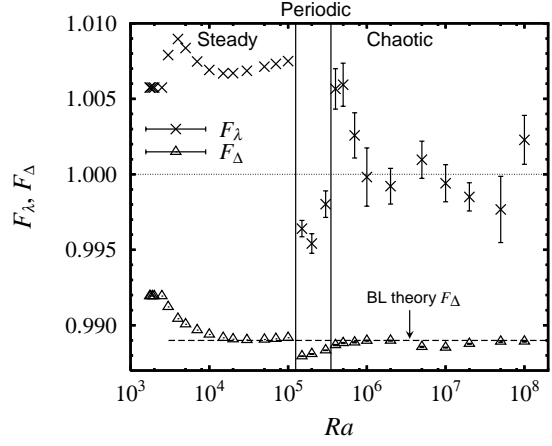,width=6cm,angle=-90}
\vspace*{0.3cm}
\end{center}
\caption{
$F_{\lambda}$ and $F_{\Delta}$ (defined in (\ref{nu_ratio}) and in the text before) versus Rayleigh number $Ra$ 
for water at fixed values for $T_m=40^o$C and NOBness $\Delta = 40$K. 
The dashed line corresponds to $F_{\Delta}$ resulting from the BL theory developed in \cite{ahl06}.
Reasonable agreement between BL theory and DNS is observed. $F_{\lambda}$ in the presently considered water case 
is compatible with $F_{\lambda} = 1$ for larger $Ra$, as in experiment. The significant $Ra$-dependence for 
smaller and medium $Ra$ seems to reflect the changes of the coherent flow structures still present at these $Ra$.
In particular one recovers the window of quite different behavior slightly above $Ra = 10^5$. 
} 
\label{f1f2-vs-ra}
\end{figure}

\begin{figure}
\begin{center}
\epsfig{file=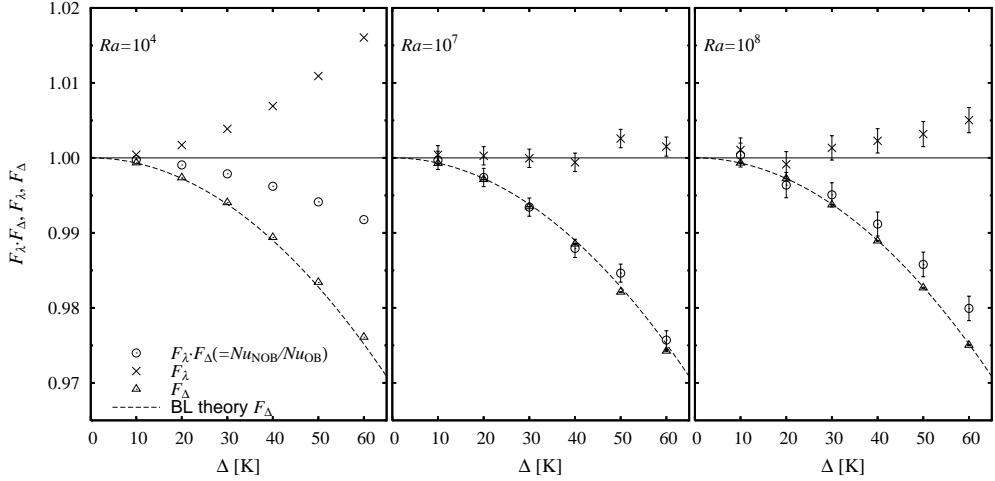,width=6.5cm,angle=-90}
\end{center}
\caption{
Nusselt number ratio 
$Nu_{NOB}/Nu_{OB} = F_{\lambda} \cdot F_{\Delta}$ together with the factors $F_{\lambda}$ and $F_{\Delta}$ 
versus NOBness $\Delta$ for fixed Rayleigh numbers (a) $Ra=10^4$, (b) $Ra=10^7$, and (c) $Ra=10^8$. The working liquid
is water at $T_m=40^o$C. The dashed lines correspond to $F_{\Delta}$ resulting from the
BL theory developed in \cite{ahl06}.
} 
\label{f-vs-del}
\end{figure}

\begin{figure}
\begin{center}
\epsfig{file=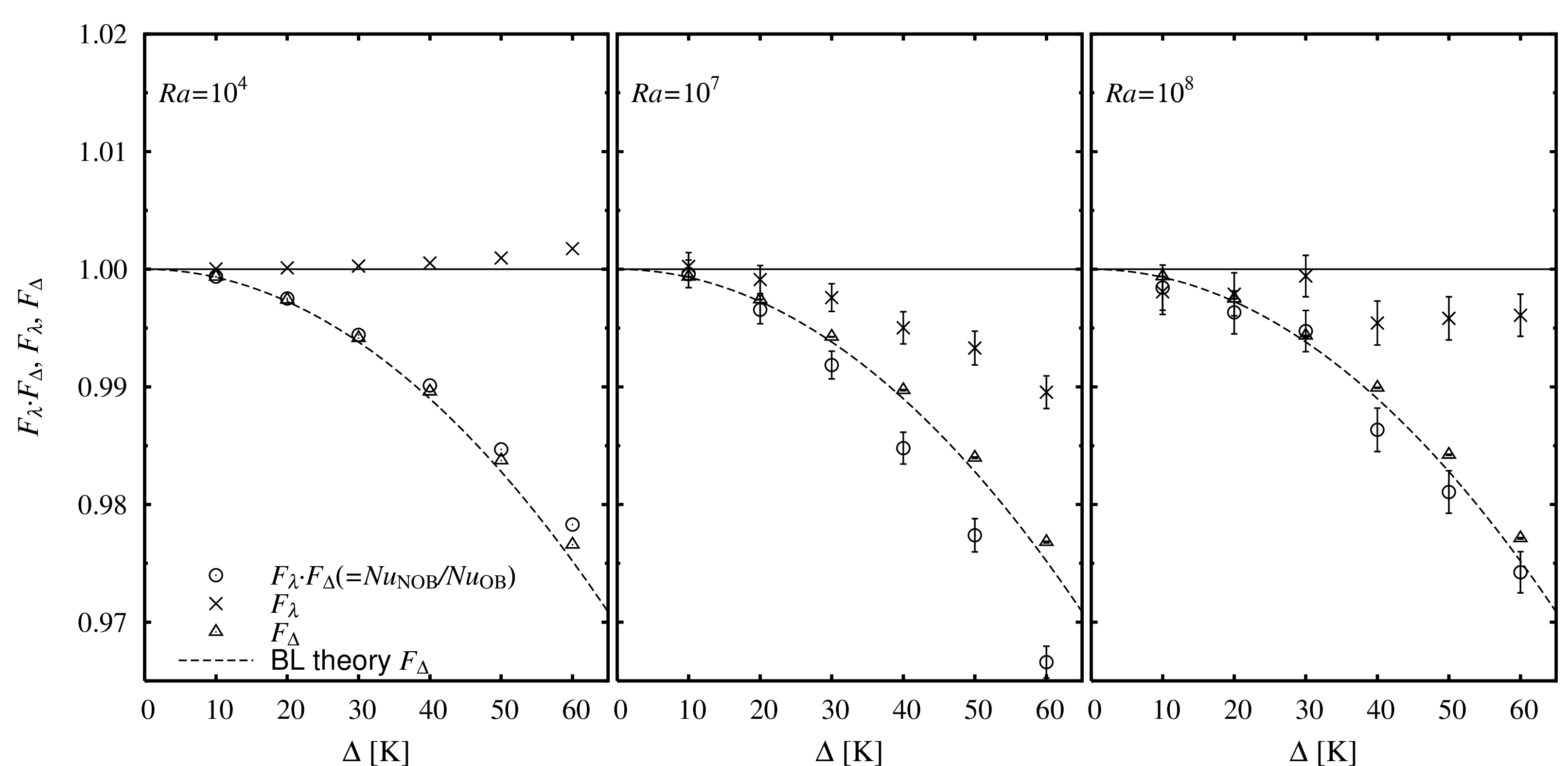,width=6.5cm,angle=-90}
\end{center}
\caption
{Same as figure \ref{f-vs-del}, but now with a temperature independent thermal expansion coefficient $\beta_m$, 
i. e., with only linear temperature dependence of $\rho(T)$ as in eq. (\ref{rho-lin}). 
The $\Delta$-dependences of the factors $F_{\Delta}$ and $F_{\lambda}$ and thus $F_{\Delta} \cdot F_{\lambda}$ are
different from those in the case of fully temperature dependent $\beta(T)$.
}
\label{f-vs-del_fixed_b}
\end{figure}

\section{Mean heat flux}\label{sec4}
Similarly to $T_c$, the Nusselt number ratio $Nu_{NOB}/Nu_{OB}$ displays an only weak dependence on $Ra$, 
see figure \ref{nu-ratio-vs-ra}. This ratio has been written in (\ref{nu_ratio}) in terms of two factors, 
$F_{\Delta}$ and $F_{\lambda}$. The latter one measures the changes of the thermal BL widths caused by NOB 
conditions. These widths of the temperature profile are sketched in figure \ref{sketch_bl}; quantitatively we define 
them in terms of the temperature slopes at the plates.
\be
\lambda_{t}^{sl} \equiv \frac{\Delta_t}{\mid \partial_z \langle T \rangle_{A_t,\tau} \mid} ,  \ \ \ 
\lambda_{b}^{sl} \equiv \frac{\Delta_b}{\mid \partial_z \langle T \rangle_{A_b,\tau} \mid } .
\label{thicknesses}
\ee
In figures \ref{f1f2-vs-ra} and \ref{f-vs-del} we reveal the origin of the Nusselt number modification in the 
NOB case. In general it is a combination of the $\Delta$-dependence of both factors $F_{\lambda}$ and $F_{\Delta}$ 
in the product (\ref{nu_ratio}). As shown in figure \ref{f1f2-vs-ra} for fixed non-Oberbeck-Boussinesqness $\Delta$, 
$F_{\Delta}$ displays a weak dependence on $Ra$ for $Ra\geq 10^4$. This can be understood 
from the weak $Ra$-dependence of the center temperature $T_c$ on the NOB changes of the material parameters 
(see figures \ref{tc-vs-del} and \ref{tc-vs-ra}), and because 
$F_{\Delta}=((\kappa_t-\kappa_b)T_c-\kappa_t T_t+\kappa_b T_b)/(\kappa_m\Delta)$ depends on $T_c$ only
(for given $T_t$ and $T_b$). 

On the other hand the factor $F_{\lambda}$, describing the variation of the 
thermal BL thicknesses, shows a rather weak but obvious dependence on the RB flow regimes. For the fully chaotic regime 
$(Ra\geq 10^6)$, the deviation of $F_{\lambda}$ from $F_{\lambda} = 1$ is much smaller than that of $F_{\Delta}$, 
which indicates that $Nu_{NOB}/Nu_{OB}$ is dominated here by $F_{\Delta}$ and thus by the behavior of the center 
temperature. As shown in figure \ref{f-vs-del}, which displays the dependences on the NOBness $\Delta$ for the 
largest analyzed $Ra=10^7$ and $10^8$, the $\Delta$-dependence of $F_{\lambda}$ happens to be very small,
and for those Rayleigh numbers $F_{\lambda} \approx 1$ happens to be a good approximation. 
This might be due to an incidental combination of the temperature dependences of the material parameters 
$\eta(T)$, $\Lambda(T)$ and $\rho(T)$ 
around 
the chosen mean temperature $T_m=40^o$C in the case of water. The experimental finding, reported in 
\cite{ahl06}, that $F_{\lambda} \approx 1$ in a similar $Ra$-range, for the same $T_m=40^o$C, 
and for $\Delta $ up to $40$K therefore can be 
considered as incidental. It is not a general property of NOB Rayleigh-B\'enard convection. We have also checked the 
influence of buoyancy on the BL widths as described by $F_{\lambda}$. If we disregard the $T$-dependence of 
$\beta(T)$, i. e. the non-linear temperature 
dependence of $\rho(T)$ in our numerical simulations and thus have a constant $\beta = \beta_m$, $F_{\lambda}$ shows 
a larger deviation from 1 at $Ra=10^7$ and happens to be closer to 1 at $Ra=10^4$, see figure \ref
{f-vs-del_fixed_b}, just opposite to the case with full $T$-dependence of $\beta(T)$. Still the nonlinear 
temperature dependence of $\rho$, i.e., the temperature dependence of $\beta$, has a relatively weak effect on 
$F_{\lambda}$ and $F_\Delta$ and therefore on the Nusselt-number modification.

The second conclusion we can draw from figures \ref{f1f2-vs-ra}, \ref{f-vs-del}, and \ref{f-vs-del_fixed_b} is that
in all cases $F_{\Delta}$ is correctly described by the BL theory given in \cite{ahl06}. As $F_{\Delta}$ 
can be calculated from $T_c$ only, this is of course to be expected, since $T_c$ 
is well described by the extended Prandtl-Blasius BL theory (see figure \ref{tc-vs-del}).

\section{The flow structure and various wind amplitudes}\label{sec5}
\begin{figure}
\begin{center}
\vspace*{0.3cm}
\epsfig{file=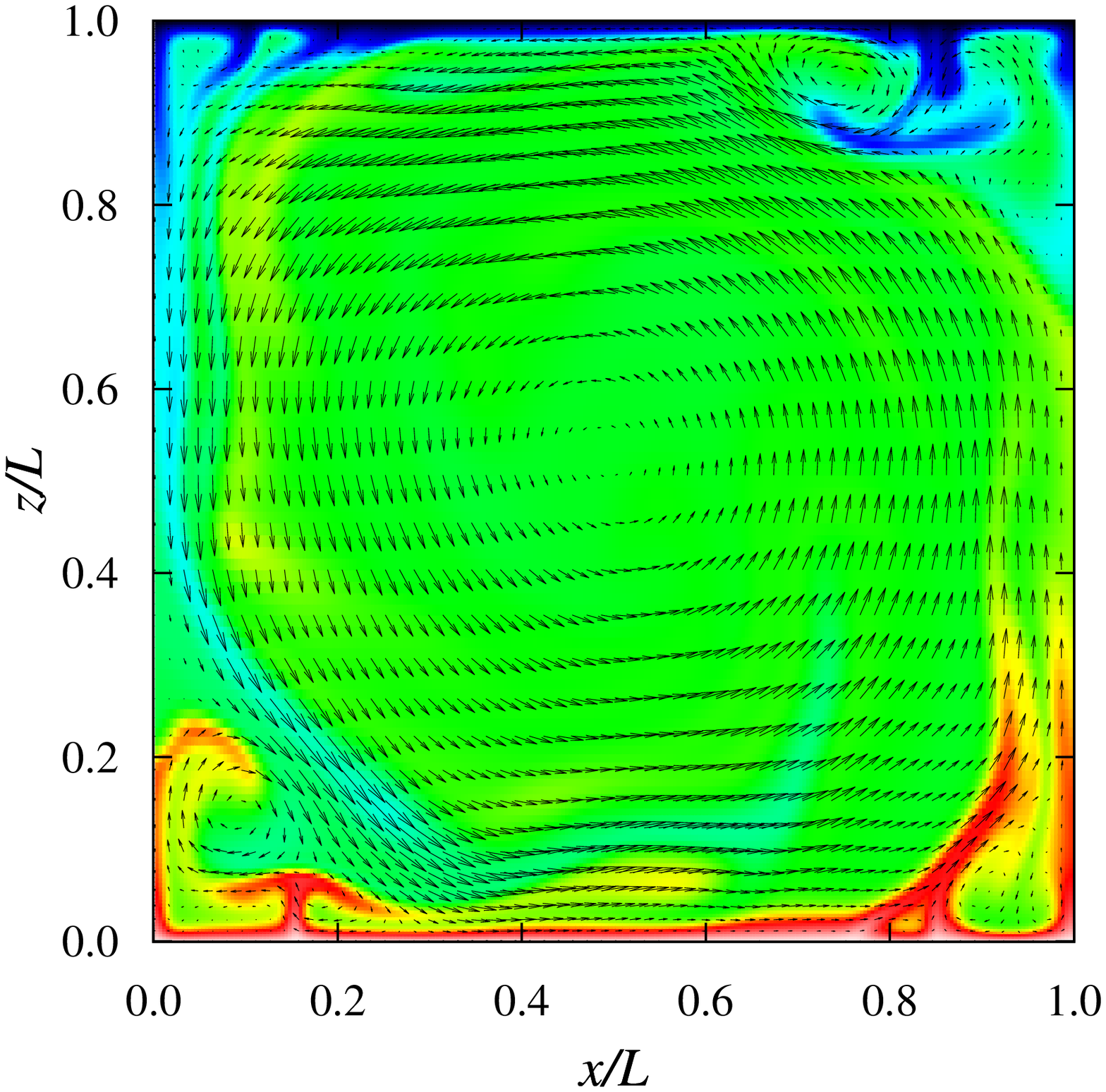,width=6.5cm,angle=-90}
\epsfig{file=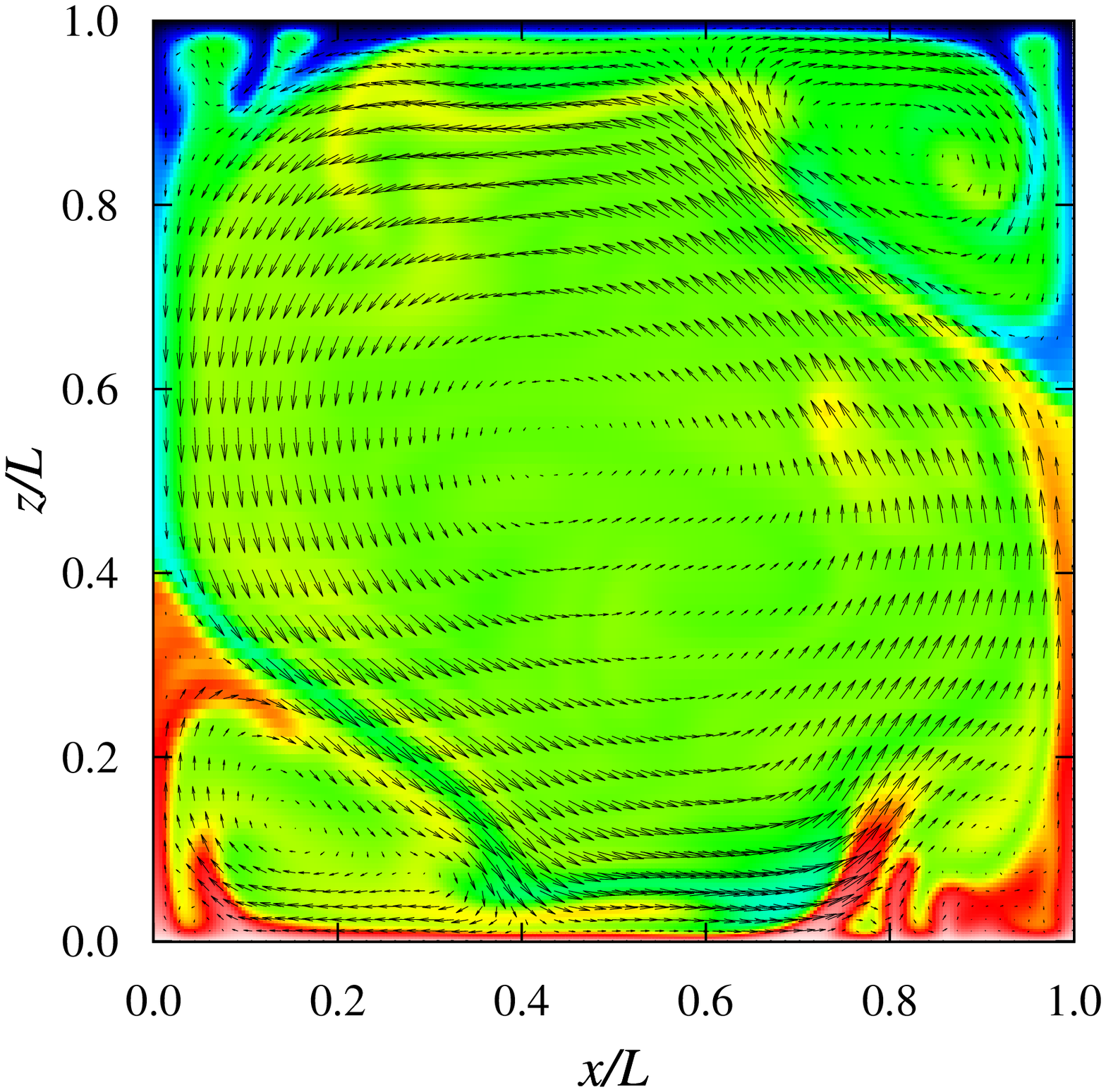,width=6.5cm,angle=-90}
\vspace*{0.3cm}
\end{center}
\begin{flushright}
\epsfig{file=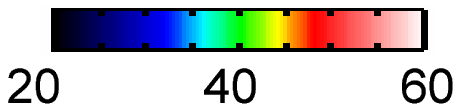,width=2.5cm,angle=0}
\end{flushright}

\caption{(color)
Snapshots of the velocity (arrows) and temperature (color) fields for $Ra=10^8$ at $T_m=40^o$C, working 
fluid water. The left panel corresponds to the OB case (all material properties are kept temperature 
independent, taken at $T_m$), the right one to the NOB case, both with the same 
$\Delta = 40$K. The temperature color scheme is in $^o$C, same in both panels.
} 
\label{snap}
\end{figure}

We focus now on the structure of the flow field or "wind" in thermal convection. Although in our 2D simulations 
we miss interesting but 
typically 3-dimensional flow modes (cf. \cite{ahl09} for a summary), even in two dimensions the velocity field is 
rather complex, as the snapshots in figure \ref{snap} display. Nevertheless such convection fields have mostly been 
described by only one single amplitude $U$.
This wind amplitude $U$ is a crucial parameter both in the general OB theory (\cite{gro00,gro01,gro02,gro04}) 
as well as in the extended BL theory dealing with NOB effects (\cite{ahl06}). The GL theory hypothesizes that one needs 
only one single mean wind amplitude to describe the heat transport (also for large $Ra$) and that this amplitude is essentially uniform throughout the cell. 

The extended BL theory, developed in \cite{ahl06} for NOB situations, assumes that such a {\it uniform} wind is 
still present even under NOB conditions  and 
that in particular the top and bottom BLs see the very same wind amplitude $U_t = U_b = U_{NOB}$,
in spite of the NOBness. The amplitude $U_{NOB}$ is allowed to be different from $U_{OB}$, but for the Nusselt number
calculations (not for $T_c$, as detailed above) its value
has to be taken as a parameter of the theory. It is this character of $U_{NOB}$ as a boundary condition for 
the Prandtl-Blasius BL equations which leaves the BL theory incomplete for calculating the heat transport across 
the RB cell. Thus neither the Reynolds nor the Nusselt number deviations under NOB conditions can be predicted, 
unless further input (data or assumptions) is introduced (as for instance $F_{\lambda} = 1$). 

We organize our analysis of the 2-dimensional flow structure as follows. Starting with the visualization 
of the dynamical flow fields, we next introduce a time averaged convective Eulerian field. 
Then we discuss several sensible possibilities to adequately 
define relevant wind amplitudes quantitatively. 
Finally, we present our results about the dependences of the $U$-amplitudes 
or $Re = U/(\nu L^{-1})$ numbers on the Rayleigh number $Ra$ and the NOBness $\Delta$. 

\subsection{Dynamical features}
\label{dynamical features}

For RB convection in water in an aspect ratio $\Gamma = 1$ container it is known from 
experiment and also noticeable in our 2D DNS flow as shown in the snapshots in Fig.\ref{snap}, that a large scale 
circulation (LSC) with an extension comparable to the box size $L$ is present both in the OB and NOB cases at 
$Ra = 10^8$. In addition and on top of this LSC there are of course fluctuations of the ${\bu},T$-fields.

The time development of the numerical ${\bu}$-field shows {\it reversals} in the circulation sign,
in agreement with experimental observations (\cite{sre02,bro05b,bro06,bro07a,ahl09})
and earlier numerical simulations (\cite{han92}). In 3-dimensional experiment
these reversals can occur either by rotation of the convection roll's plane 
or by cessation and restart; in the 2-dimensional numerics of course only the latter type of reversal occurs. 
Several models have been developed for these reversals, see e.g.\ \cite{sre02,fon05,ben05,bro07a}.
We shall report on details of our results about the statistical properties of the reversals in our
2D simulations elsewhere. In the context of the present paper these reversals only complicate the statistical analysis 
of the flow field, as long-time averages of the velocity field become zero and wash out the flow structures.

How then to obtain the main features of the dynamical, complex time-dependent ${\bu}$-field? To achieve them we 
consider conditionally time averaged Eulerian fields as well as several profiles, which partly take the 
fluctuations into account too. In addition a global, energy based wind amplitude is introduced.

\subsection{Conditionally time averaged velocity fields}
\label{conditional averages}
To overcome the problem that long time averages due to the statistical flow reversals give zero 
velocity everywhere, we perform
{\it conditioned} time averages, which take the time dependent rotational direction of the wind 
into account. This instantaneous rotational direction is identified by the sign of the vorticity at the center of 
the box. Whenever the wind reverts its direction, before performing the standard time-averaging the velocity field 
is mirrored along the vertical center-line. Respecting this, from the full velocity field ${\bm u}({\bm x},t)$ we can 
compute another, time averaged, complete 2D Eulerian-type velocity field $\overline{{\bm u}}({\bm x})$, in which 
the respective local direction of the velocity is coupled to the sign of the central vortex. Component-wise we 
define this conditionally time averaged flow field as
\begin{equation}
\begin{split}
\overline{u_x}(x,z)
&= -\bigT^{-1}\int_{t_0}^{t_0+\bigT}\!\!\!{\rm d}t\ 
 u_x(\tilde{x},z,t)\ \mbox{sign} ~\omega_c(t),\\
\overline{u_z}(x,z)
&=\ \bigT^{-1}\int_{t_0}^{t_0+\bigT}\!\!\!{\rm d}t\ 
u_z(\tilde{x},z,t).
\label{eq:ctv_uw}
\end{split}
\end{equation}
Here  $\bigT$ denotes the averaging time, sign$ ~\omega_c (t)$ is the sign of the vorticity 
$\omega(x,z) \equiv \partial_z u_x -\partial_x u_z$ at the center of the box $(x,z) = (\tfrac{L}{2}, \tfrac{L}{2})$, 
and $\tilde{x}$ is  
$$
\tilde{x} = L/2 + (L/2 -x) \cdot \mbox{sign}~\omega_c (t) \ . 
$$ 
In a similar manner we also define the conditionally averaged velocity squares as
\begin{equation}
\begin{split}
\overline{u_x^2}(x,z)
&= \bigT^{-1}\int_{t_0}^{t_0+\bigT}\!\!\!{\rm d}t\ 
u_x^2(\tilde{x},z,t),\\
\overline{u_z^2}(x,z)
&=\ \bigT^{-1}\int_{t_0}^{t_0+\bigT}\!\!\!{\rm d}t\ 
u_z^2(\tilde{x},z,t).
\label{eq:ctv_uw2}
\end{split}
\end{equation}
Rather than conditioning on the sign of the vorticity at the center, one could also condition on the sign of 
the total angular momentum, as has been done by \cite{hei06} (which minimizes the contributions of some 
high-frequency oscillations), 
but for the purpose of this paper the difference between these two types of conditional averaging have 
turned out not to be relevant.

\begin{figure}
\begin{center}
\epsfig{file=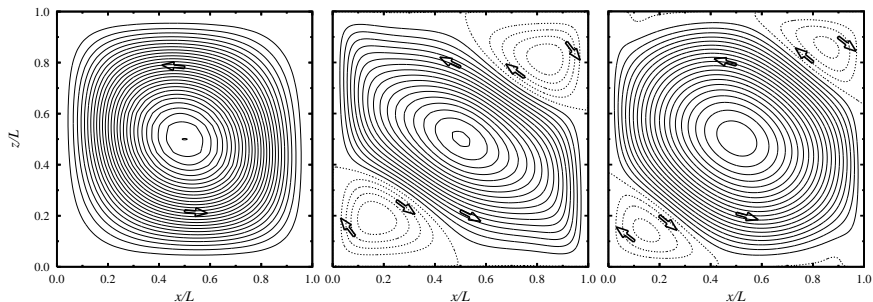,width=4cm,angle=-90}
\end{center}
\caption{Lines of constant values for the conditionally time averaged velocity field $\overline{\bm u}(x,z)$ 
at different Rayleigh numbers $Ra= 10^4, 10^6, 10^8$ in OB convection. Counterclockwise velocity direction 
is drawn with solid lines, while the clockwise ones are indicated by dotted lines. The OB flow structure, 
which already develops secondary (counter) rolls in the corners, enjoys top-bottom symmetry.   
} 
\label{snap_cs}
\end{figure}

Conditionally time averaged fields $\overline{{\bm u}}({\bm x})$, being time independent objects, allow to 
visualize the persistent spatial structures in the flow field. Some conditionally time averaged fields obtained 
for different $Ra$ numbers are shown in Fig.\ref{snap_cs}. 

Note again the large circulation roll in the center range but also the secondary counter-rotating rolls in the 
corners. We remark that such secondary circulation rolls in nearly-2D convection have been experimentally detected 
by \cite{xia03}. In our simulations secondary rolls appear for $Ra > 10^5$. We interpret the secondary rolls as 
caused by boundary layer separations, which are known to occur when a flow is heading a perpendicular wall. 
They might be 
considered as kind of "wakes" behind the separation. Remarkably, the secondary roles are of considerable size 
for the BL separations of the up and down going flows, which approach the top and bottom plates, 
respectively, but are nearly invisible for the horizontal ones, which approach the side walls (at least for these 
Rayleigh numbers). This might be attributed to the plume creations in the BLs on the bottom and top plates, which 
affect the horizontal but not the vertical sections of the flow. 

We emphasize that the convergence of the statistics implies that a center-point symmetry should be established 
for the conditionally time averaged field $\overline{{\bm u}}$. Indeed, within 5$\%$ precison we achieve such a 
center-point symmetry in our numerical simulations. As we shall see later this accuracy is by far sufficient to 
discriminate the main NOB effects in comparison to the OB results.

\begin{figure}
\begin{center}
\epsfig{file=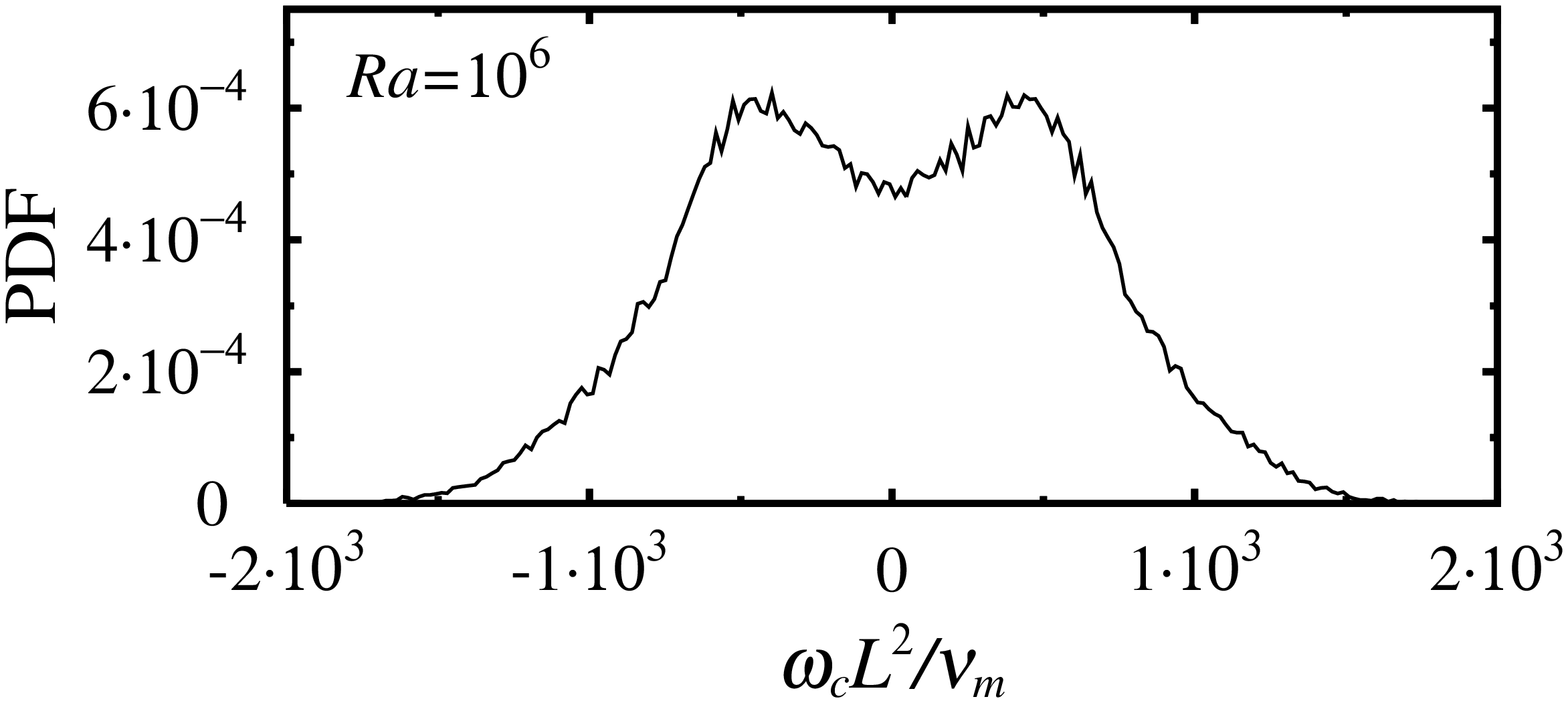,width=4cm,angle=-90}
\epsfig{file=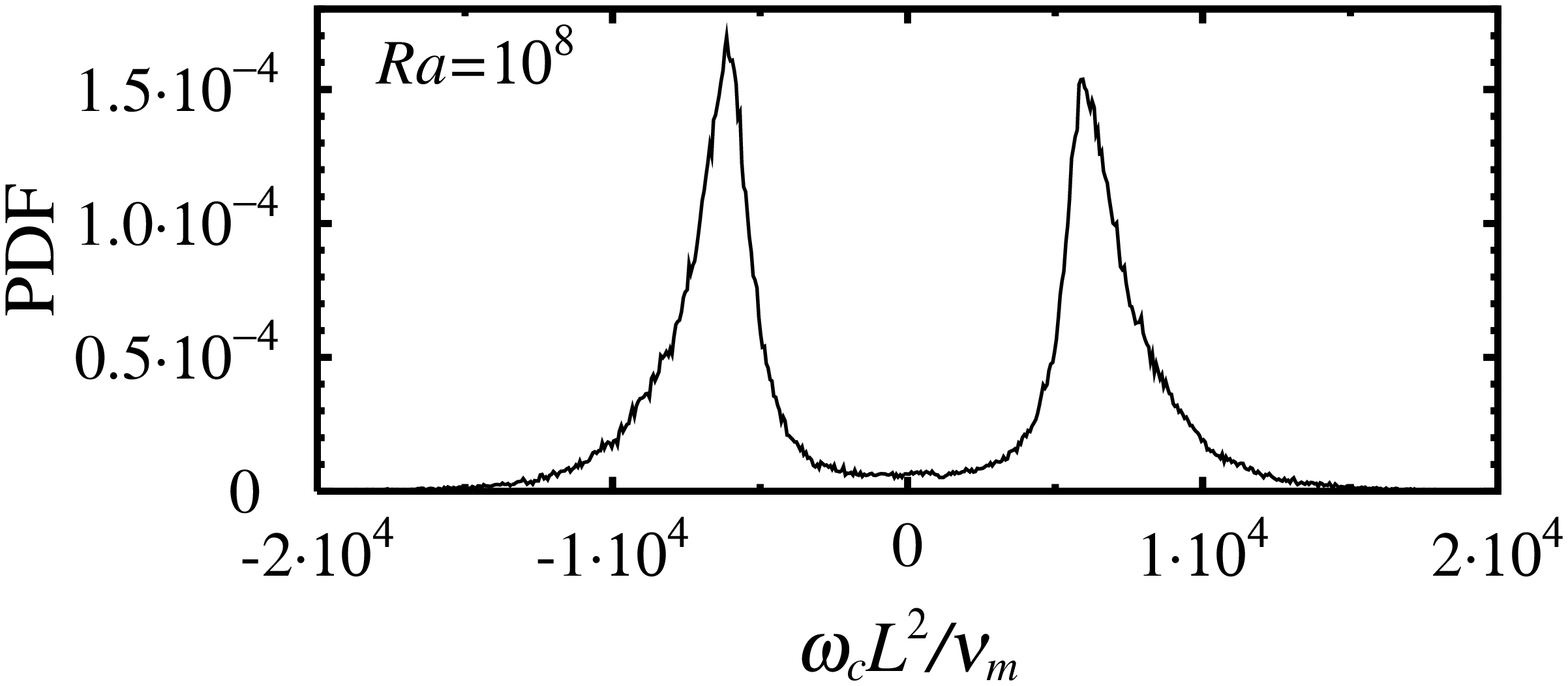,width=4cm,angle=-90}
\end{center}
\caption
{Probability density function of the vorticity $\omega_c$ at the cell center at $Ra = 10^6, 10^8$.
} 
\label{pdfvort}
\end{figure}

For information we add the probability density function (PDF) of the center vorticity $\omega_c$,  
(nondimensionalized by the molecular vorticity $\nu_m L^{-2}$), see Fig.\ref{pdfvort}. Already for $Ra = 10^6$ 
two preferred values of the LSC can be recognized, reflecting clockwise and counter-clockwise 
rotation of the large scale 
convection roll. For $Ra=10^8$ the two preferred vorticities are even more pronounced,
reflected by the sharp peaks in the PDF. While in the chaotic phase ($Ra \approx 10^6$) the PDF is still broad, 
the flow cessations at $Ra \approx 10^8$ are very fast events, since the small probability for $\omega_c = 0$ 
suggests that the flow changes its direction more or less momentarily, leaving only a very small 
probability to find $\omega_c = 0$.  

\subsection{Wind profiles: amplitudes}
\label{wind profiles: amplitudes} 
Because even the conditionally time-averaged velocity field as shown in Fig.\ \ref{snap_cs} has a 
rich spatial structure, it is not immediately apparent how to define ``a characteristic single wind 
velocity amplitude $U$" unambiguously. Several definitions 
have been proposed in the literature. In the present section we introduce various possible definitions for wind 
amplitudes $U$ and corresponding Reynolds numbers $Re = U L / \nu_m$, by investigating global measures as well as 
local ones, and also certain profiles.
 
i. Global. A global measure for the strength (amplitude) of the convection can be based on the volume 
average. While the velocity average over the full volume is of course zero, we can consider the velocity 
{\it rms} average and obtain the energy based wind amplitude   
\begin{equation} \label{energyamplitude}
U^{E} \equiv
\sqrt{ \left\langle ~\frac{1}{2} ~( u_x^2+u_z^2 ) ~\right\rangle_{V,t}} ~
= \sqrt{ \left\langle ~\frac{1}{2} ~( \overline{u_x^2} + \overline{u_z^2} ) ~\right\rangle_{V}}.
\end{equation}
Apparently, both the primary center roll as well as the secondary rolls contribute to the value of $U^{E}$.
While the primary roll covers the whole interior, which is expected to be mixed by turbulence and therefore has 
essentially uniform temperature $T_c$, the secondary rolls experience either the cooler top or the hotter 
bottom regions only. Since under NOB conditions the viscosity deviations are different near the top and bottom 
(and, of course, from the bulk), 
the secondary rolls are expected to have different properties among each other as well as relative to the 
primary center roll. $U^{E}$ represents a well defined mixture of all of them.  

ii. Local. To deal with the top and bottom differences, local wind amplitudes may be introduced by considering 
the values of the conditionally time averaged field $\overline{{\bm u}}({\bm x})$ at particular spatial 
positions ${\bm x}_j$ in the flow field. Particular positions are e.g. those, where the conditionally 
time-averaged velocity field 
$\overline{{\bm u}}$ has peak values $U^{P_j}$ along some vertical lines, labelled by $j$. In the following we shall 
consider such local peak amplitudes $U^{P_j}$ at peak positions on vertical lines with abscissas 
$x_j = \tfrac{3L}{4}, \tfrac{L}{2}$, and $\tfrac{L}{8}$. These points $P_j$ roughly correspond to the positions 
where the main, primary circulation roll is strongest ($x_1=\tfrac{3L}{4}$), to the flow maxima along 
the center line of the container 
($x_2 =\tfrac{L}{2}$), and to the region where the counter-rotating roll is well developed ($x_3 =\tfrac{L}{8}$). 

iii. Profiles. Mixed type wind amplitudes, neither fully global nor fully local, can be introduced as area 
averages or rather, in 2D, as line averages. The areas (lines) can either be chosen as top/bottom plate parallel or 
as side wall parallel. The corresponding area averaged wind amplitudes then either, as in the case of plate parallel 
averaging, depend on the height $z$ of the area (line) and lead to $z$-dependent wind profiles; this is relevant for 
the horizontal wind and its vertical profile $U_x(z)$. Or they depend, as in the case of vertical line averaging, on 
the $x$-distance to the side walls and lead to $x$-dependent profiles; these are relevant for the up rising or down 
falling flow and its horizontal profile, denoted as $U_z(x)$.  

Such area (line) and time averages are the relevant quantities in the well known relations between the dissipation 
rates $\varepsilon_u$, $\varepsilon_{\theta}$, and the nondimensionalized heat current density, i.e., the Nusselt 
number $Nu$. While the dissipation rates are volume averages, the Nusselt number is defined in terms of 
a horizontal area average, 
which by conservation of energy even is independent of height $z$. One contribution to $Nu$ comes from the 
$\langle T u_z \rangle_{A,t}$-correlation, the other one from the gradient of the temperature profile 
$\langle T \rangle_{A,t}(z)$. This justifies to introduce the mentioned area (line) averages also for velocity 
components. In particular, the vertical profile of the horizontal velocity is of interest, $U_x(z)$, as well as the 
horizontal profile of the vertical velocity, $U_z(x)$, precisely defined as 
\begin{equation} \label{area-averages}
U_x(z) = \left\langle \overline{u_x} \right\rangle_{x(z)}(z), \hspace{8mm} \mbox{and} \hspace{8mm}  
U_z(x) = \left\langle \overline{u_z} \right\rangle_{z(x)}(x).
\end{equation}
where $\langle\ldots\rangle_{x(z)}$ and $\langle\ldots\rangle_{z(x)}$ represent the line averaging along 
the $x$ direction for fixed $z$ or along the $z$ direction for fixed $x$. 

\begin{figure}
\begin{center}
\epsfig{file=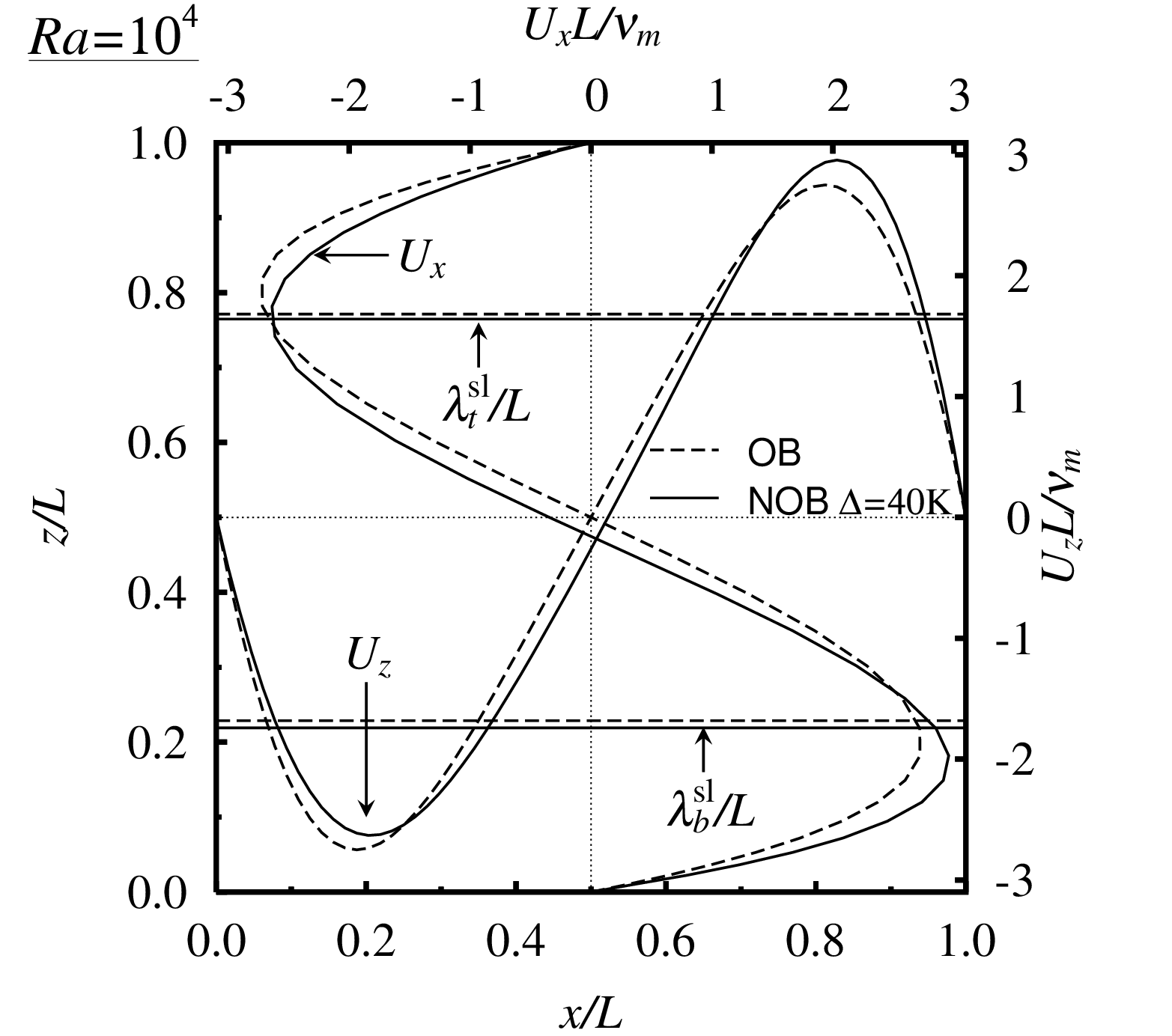,width=5.6cm,angle=-90}
\epsfig{file=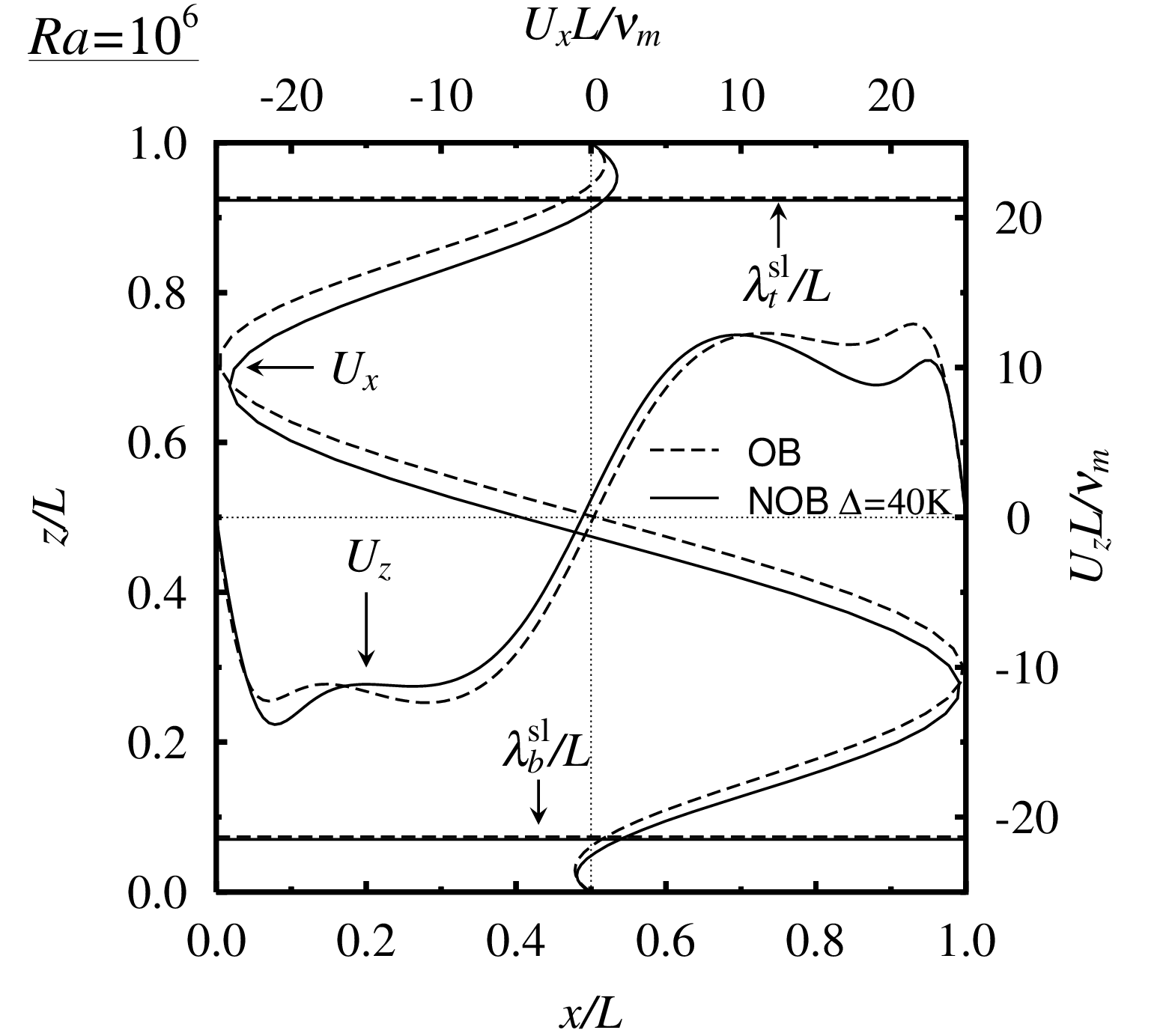,width=5.6cm,angle=-90}
\epsfig{file=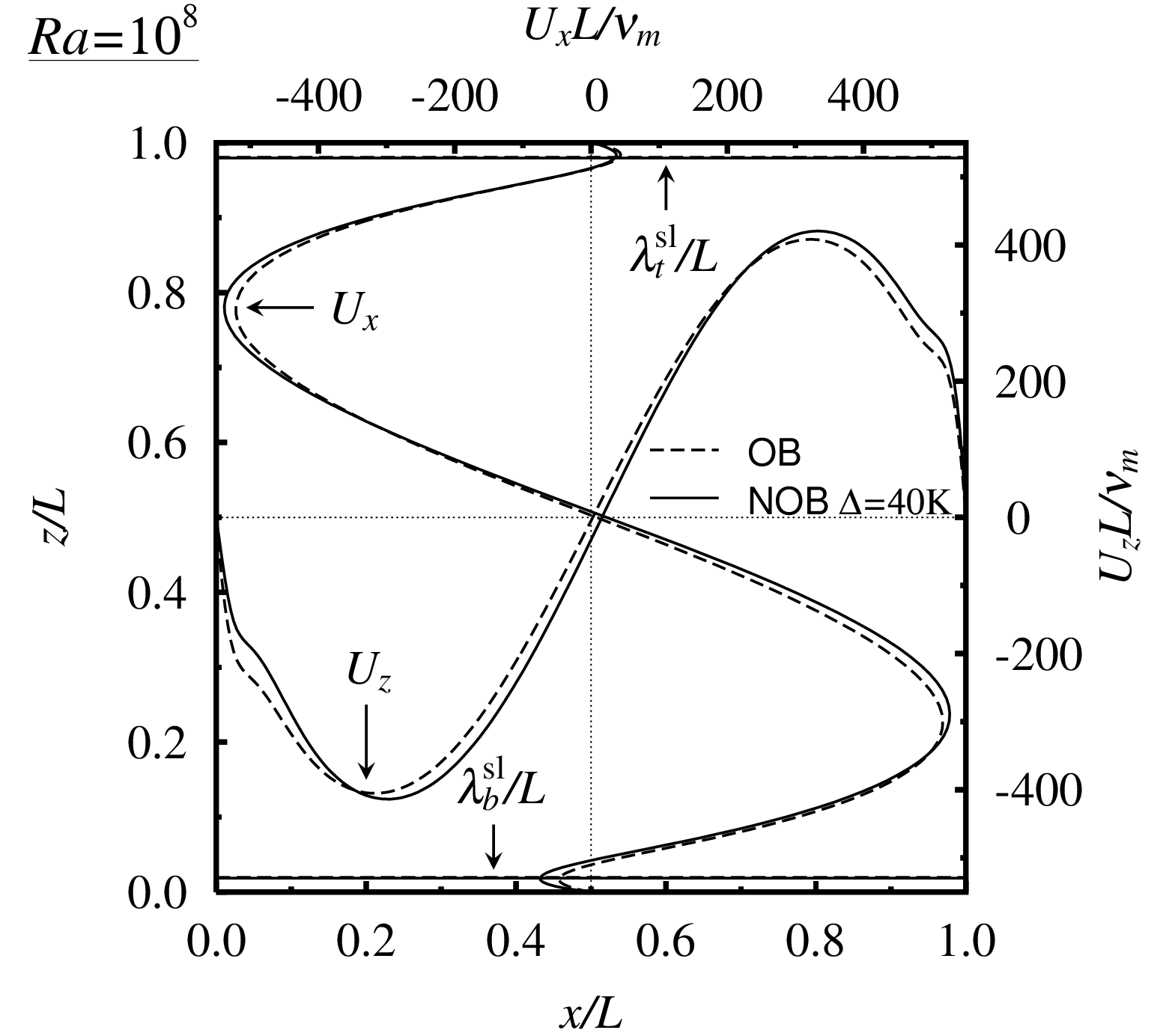,width=5.6cm,angle=-90}
\end{center}
\caption{ 
The vertical and horizontal velocity profiles $U_x(z)$ and $U_z(x)$ as derived from 
the conditionally time averaged velocity components $\overline{u_x}$ and $\overline{u_z}$ at three different $Ra$ 
numbers $10^4,10^6,10^8$. The abscissa and ordinate scales are the dimensionless 
width and height. The upper scales show the 
horizontal velocity profiles nondimensionalized with $\nu_m/L$, i.e., $Re_x(z)$; the right scales show the vertical 
velocity's $U_z(x)$ profiles, also nondimensionalized by $\nu_m/L$. Note the increase of the $Re_{x,z}$ scales 
with increasing $Ra$. Dashed lines indicate the OB case, full lines NOB case. In both cases $\Delta = 40$K. Also 
indicated are the corresponding thermal slope BL widths $\lambda_{b,t}^{sl}$, which strongly decrease with $Ra$.
}
\label{ucond_profiles}
\end{figure}

\begin{figure}
\begin{center}
\epsfig{file=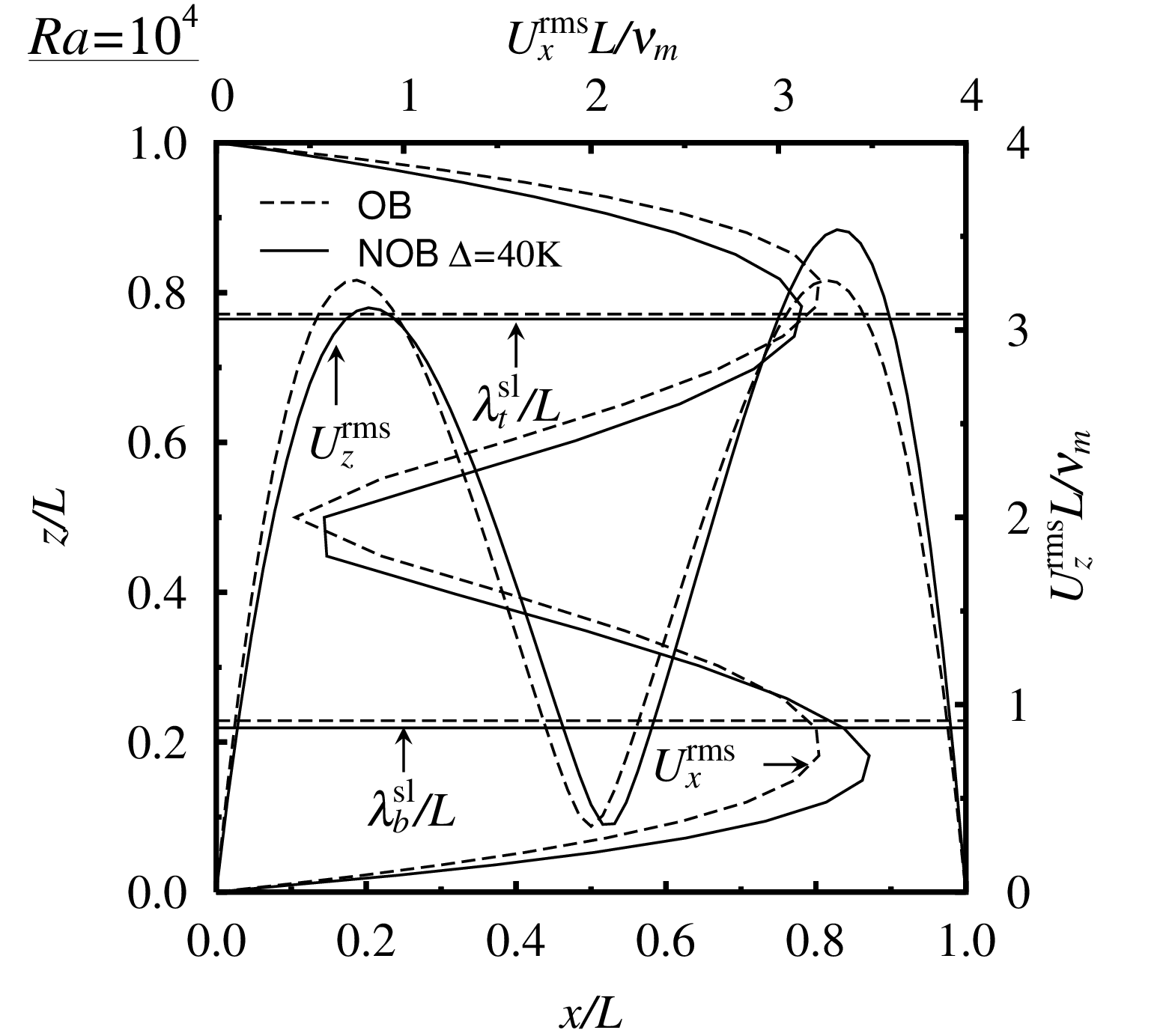,width=5.6cm,angle=-90}
\epsfig{file=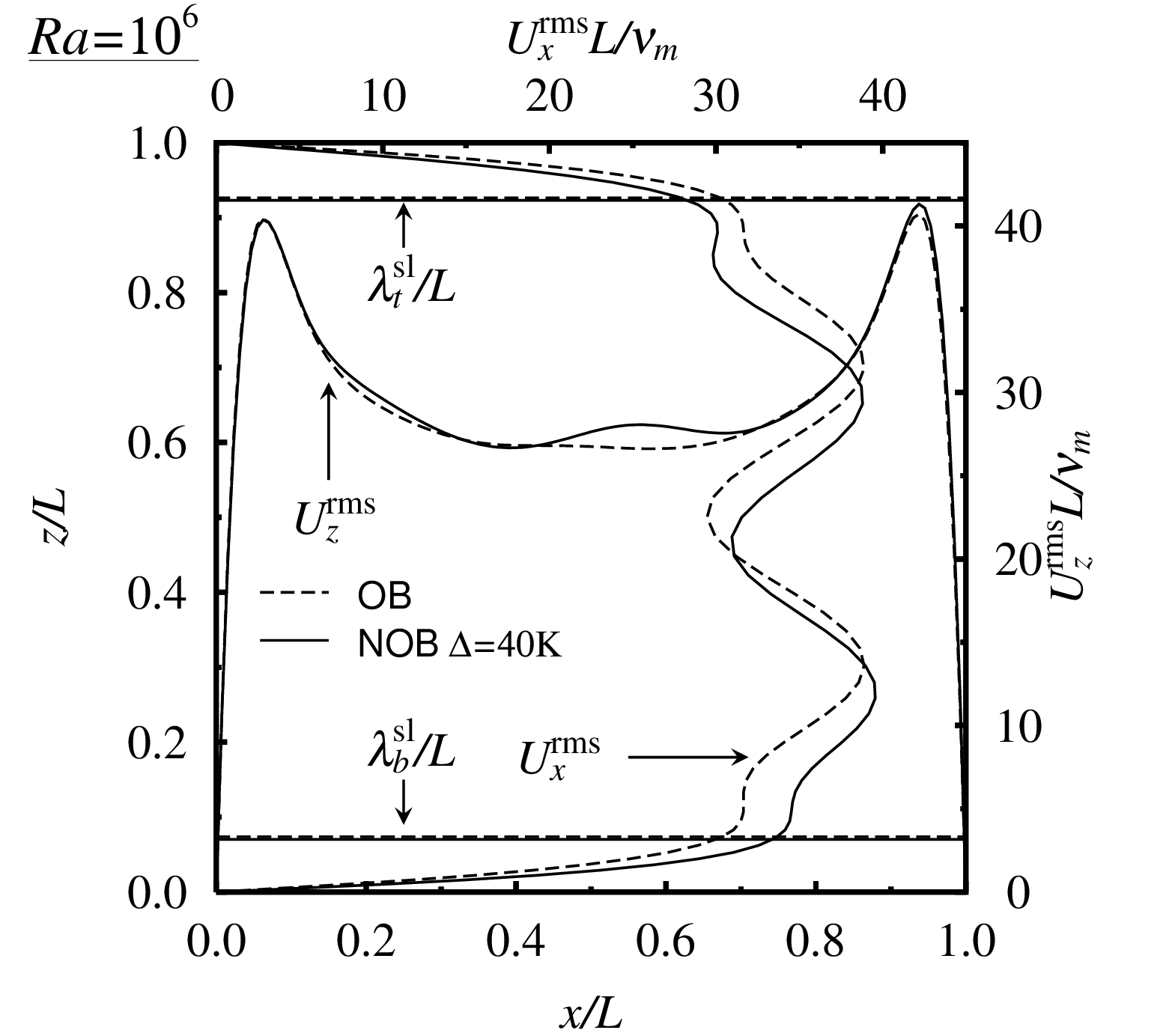,width=5.6cm,angle=-90}
\epsfig{file=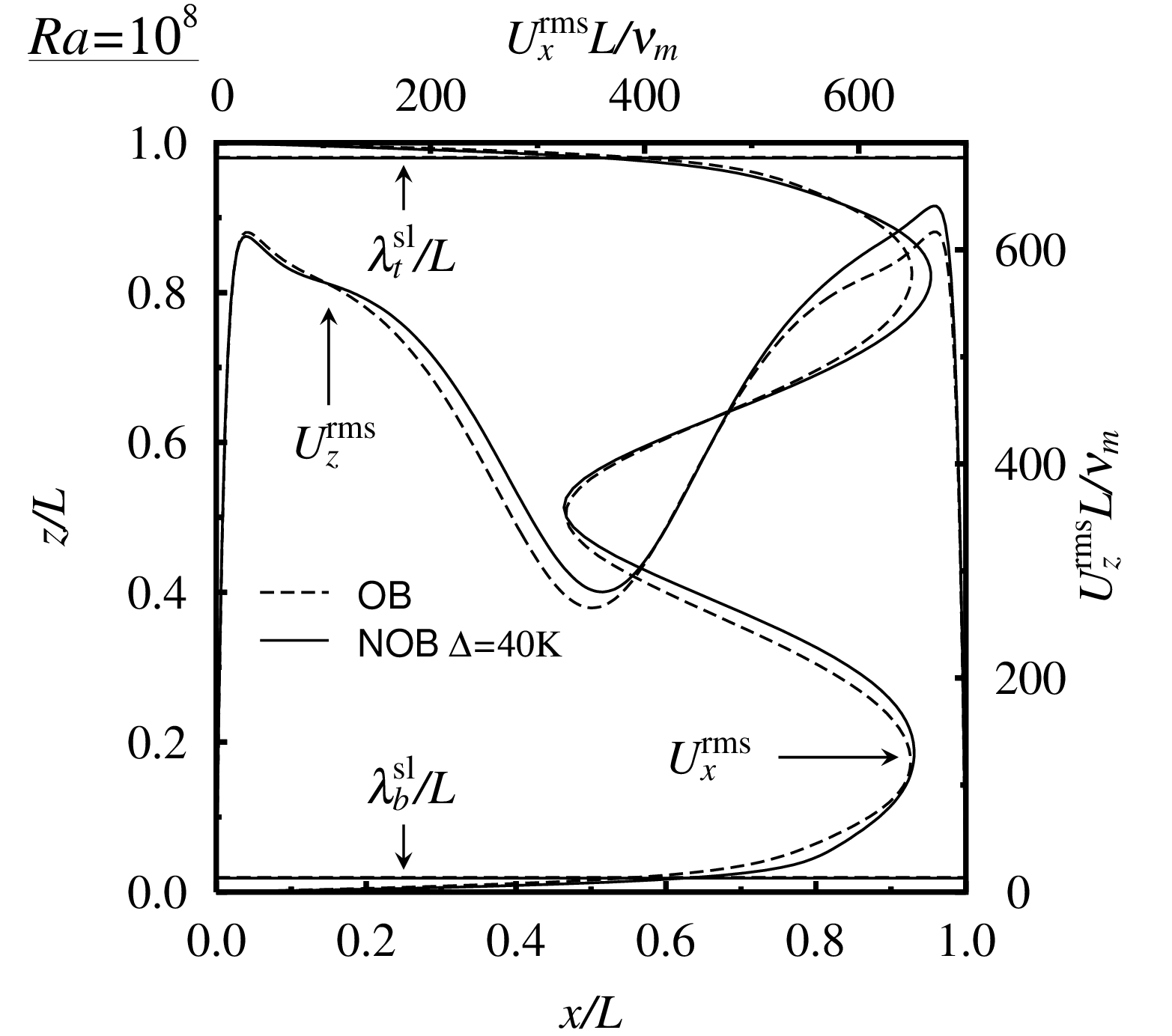,width=5.6cm,angle=-90}
\end{center}
\caption{ 
The vertical and horizontal rms velocity profiles 
$U_x^{rms}(z)$ and $U_z^{rms}(x)$
as derived from the area averaged rms fields 
$\sqrt{\langle u_x^2 \rangle_{x,t}}(z)$ and 
$\sqrt{\langle u_z^2 \rangle_{z,t}}(x)$ 
at three different $Ra$ numbers $10^4,10^6,10^8$.
The scaling schemes for the abscissa and ordinate,
the representations of the lines, 
and the conditions are same as figure \ref{ucond_profiles}. Note the different magnitudes (scales) 
for the different $Ra$.
} 
\label{urmscond_profiles}
\end{figure}

Area averaged profiles are displayed in figure \ref{ucond_profiles}. Surprisingly there are ranges with 
negative (positive) $U_x(z)$ of the area averaged horizontal velocity in the immediate vicinity of the bottom (top) 
plates. Formally their origin is that the area averaged profiles $U_x(z)$ take notice of the sign of the 
corresponding local 
velocities and these in the secondary rolls are opposite to the center roll. Thus one clearly sees the effects of 
the secondary rolls in the inversion of the vertical 
profile in the neighborhood of the bottom and top plates. Physically this means that plumes in this range of small 
distances from the plates are mostly advected in the opposite direction until they come farther away.   
One also observes broken top-down symmetry. This is caused by the secondary rolls, since these 
cover regions of different temperatures. 

iv. rms profiles. In addition to area averages of the (conditionally time averaged) velocity components 
themselves one might also wish to analyze area averages of the rms fields of the corresponding components, defined as 
\begin{equation} \label{rms area-averages}
U_x^{rms}(z) = \sqrt{\left\langle \overline{u_x^2} \right\rangle_{x(z)}} ~(z)
=\sqrt{\left\langle u_x^2 \right\rangle_{x(z),t}} ~(z), 
\ \ \ \mbox{and}\ \ \ 
U_z^{rms}(x) = \sqrt{\left\langle \overline{u_z^2} \right\rangle_{z(x)}} ~(x).
\end{equation}                
These are displayed in figure \ref{urmscond_profiles}. 

In the rms profiles $U_{x,z}^{rms}$ one also observes the kink and the broken symmetry as in the averaged velocity
profiles $U_{x,z}$ in figure \ref{ucond_profiles}, but no change in sign, of course. The preferentially strong 
plume convection near the side walls is clearly reflected in the $U_{z}^{rms}(x)$ profiles and their sharper peaks 
as well as apparently significantly smaller boundary layers than in the $U_x^{rms}(z)$ profiles. 
In the vertical profiles of $U_{x}^{rms}(z)$ the near-wall inversion (owing to the opposite sign of the velocity 
in the secondary and center rolls) is not present, in contrast to the amplitude averaged $U_{x}(z)$ profile in 
figure \ref{ucond_profiles}, since $U_{x}^{rms}$ is positive everywhere. 
Instead, one recognizes the signature of the secondary roll effect as
the steep slope in the $U_{x}^{rms}$ profile between the plate and its maximum position. 

Experimentally often Lagrangian flow properties as e.g. the plume turnover times are used to characterize in 
particular the large scale coherent flow (LSC).
We emphasize that such features should be identified in the $U_x(z), U_z(x)$ profiles. The 
$U_x^{rms}(z), U_z^{rms}(x)$ profiles, instead, reflect the energy strengths of the considered components. 
In the following we will use two different amplitude velocities defined on the local maxima of  horizontal profiles $U_x(z)$ and $U_x^{rms}(z)$, denoted respectively as $U_x^{M_2}$ , $U_x^{rms,M_{rms}}$,
where the super-script  $M_2$ indicates the the second (and positive) maximum on $U_x(z)$.
In Figures \ref{ucond_profiles} and \ref{urmscond_profiles} the $z$-distances, at which the velocities 
have the values $U_x^{M_2}$ and $U_{x}^{rms,M_{rms}}$, respectively, 
are comparable. Both are located within the thermal BL at $Ra=10^4$ while they are 
outside at $Ra=10^6$ and  $10^8$. Furthermore, NOB effects on the maximum positions are Rayleigh number independent, in the sense that
one observes shifts of comparable size towards the bottom plate at $Ra=10^4$ and $ 10^6$.
However, as we shall see later, one notices slight differences because the nontrivial spatio-temporal flow structure  is differently reflected in the $U_x$ and $U_x^{rms}$ profiles.

\subsection{Scaling of amplitudes with $Ra$, Oberbeck-Boussinesq case}
\label{amplitude-scaling-OB}
\begin{figure}
\begin{center}
\epsfig{file=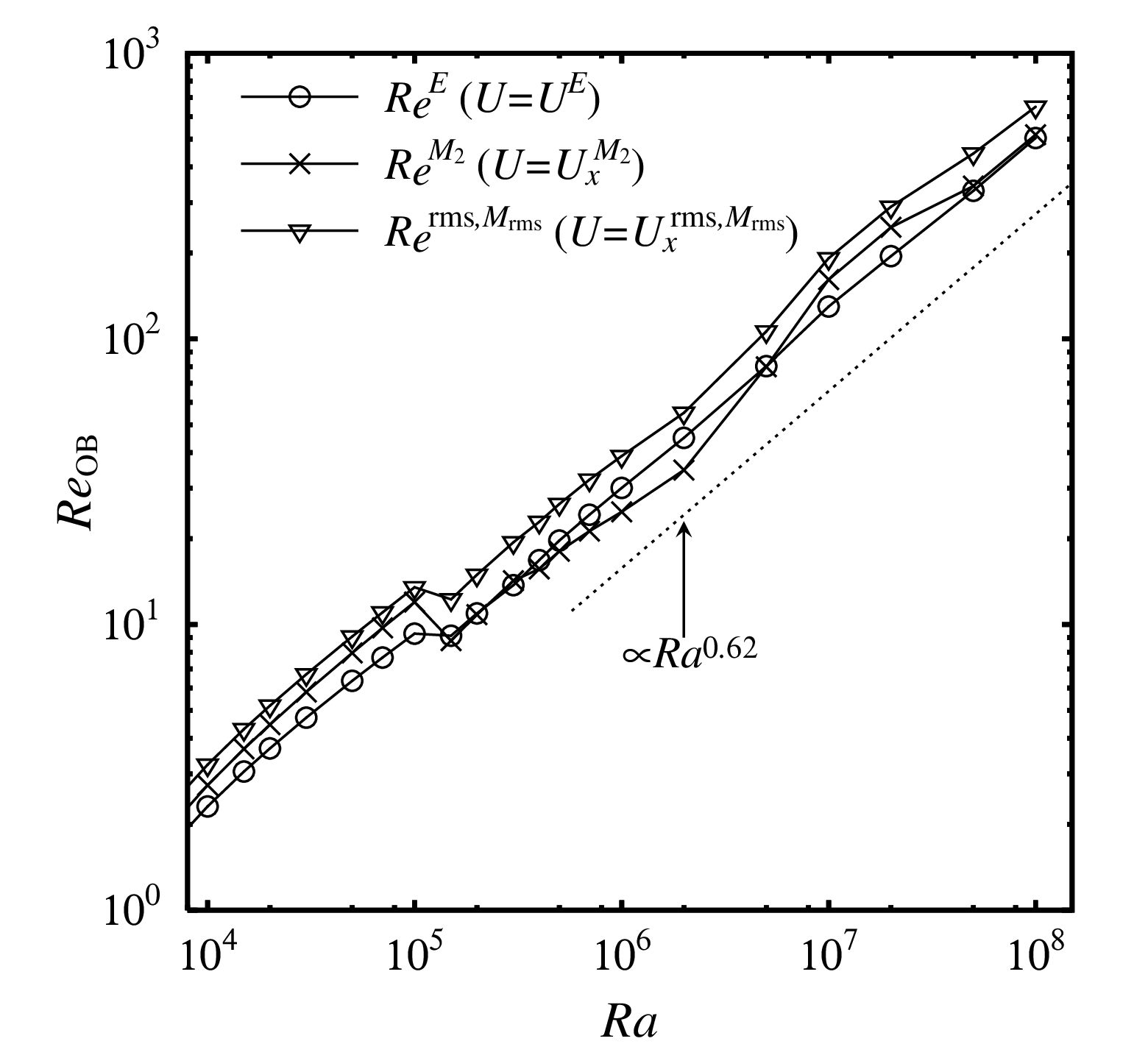,width=8cm,angle=-90}
\end{center}
\caption{
Several energy and peak based Reynolds numbers in the OB case versus $Ra$ for  water at $T_m=40^o$C. The temperature 
difference between bottom and top is $\Delta = 40$K. 
} 
\label{scale_reob_ra}
\end{figure}

Having described the flow structures and the definitions of several relevant measures for the magnitude of 
the thermally driven convection, we now offer our results on the $Ra$- and $\Delta$-dependence of the 
various $U$ amplitudes and the corresponding Reynolds numbers $Re = U / (\nu_m L^{-1})$. We start with the OB case, 
i.e., having temperature independent material parameters throughout the container, their values 
taken at the given arithmetic mean temperature $T_m$. Fig.\ref{scale_reob_ra} shows the scaling of $Re^E$, $Re^{M_2}$ and $Re^{rms,M_{rms}}$ with $Ra$ for the OB case. 

The structure in the $Re^E_{OB}$ versus $Ra$ curve around $Ra \approx 10^5$ is due to changes in the still 
present coherent flow patterns, implied by the boundary conditions. They are observed in the $Nu$ behavior too 
(not shown here), and are also detected in experiment, see for instance \cite{thre75}. For water as the working 
fluid the typical spatial coherence length (in terms of $L$) of the coherent flow structures has decreased to about 
order $0.1$ only in the $Ra$-range between some $10^7$ and $10^8$ (\cite{sug07b}). A power-law fit, $Re^E = c Ra^{\gamma}$, 
in the range $7\cdot 10^5\leq Ra\leq 10^8$, corresponding to the transition range from chaotic to turbulent 
behavior, gives the exponent $\gamma=0.616$.    

\begin{figure}
\begin{center}
\epsfig{file=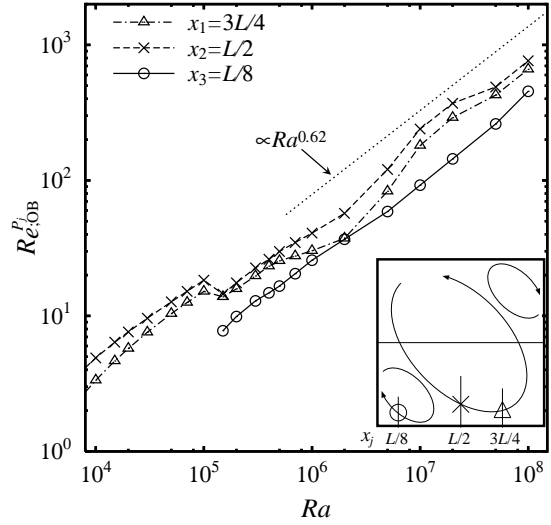,width=8cm,angle=0}
\end{center}
\caption{
Reynolds numbers $Re^{P_j}$ at $x_j = \tfrac{3L}{4} (triangles), \tfrac{L}{2} (crosses), \tfrac{L}{8}$ (circles) in the OB case
versus $Ra$ for water at $T_m=40^o$C; again $\Delta = 40$K.   The inset shows a sketch of the flow with 
the positions of the velocity peaks being indicated by the three different symbols.
} 
\label{scale_reob_cond_ra}
\end{figure}
 
The behavior of the $Re^{P_j}$ at $x_j= \tfrac{L}{8}, \tfrac{L}{2}, \tfrac{3L}{4}$, see Fig.\ \ref{scale_reob_cond_ra},
is more noisy but scaling-wise similar to (at least compatible with) $Re^E$, although in $Re^{P_j}$ the spatial 
structures of the flow field are well taken into account. $Re^E$ is more robust with 
respect to the convergence of the statistics. Experimentally at moderate $Ra$ the Reynolds numbers may scale 
differently, cf. \cite{lam02}. Here the Reynolds number based on the maximum horizontal velocity near the bottom 
plate outside the BL is reported to scale $\propto Ra^{0.70}$ for $Ra\leq 2\cdot 10^7$ and $\propto Ra^{0.495}$ for 
$Ra\geq 2\cdot 10^7$. The set of exponents $\gamma$ obtained in our simulation for $Re^E$  and the three $Re_{x_j}^P$
($\gamma \simeq 0.62$) are therefore reasonably consistent with experimental findings.

\subsection{Scaling of amplitudes with $Ra$ and $\Delta$, non-Oberbeck-Boussinesq case}
\label{amplitude_scaling_nob}

\begin{figure}

\vspace*{0.5cm}

\begin{center}
\vspace*{-2em}
\epsfig{file=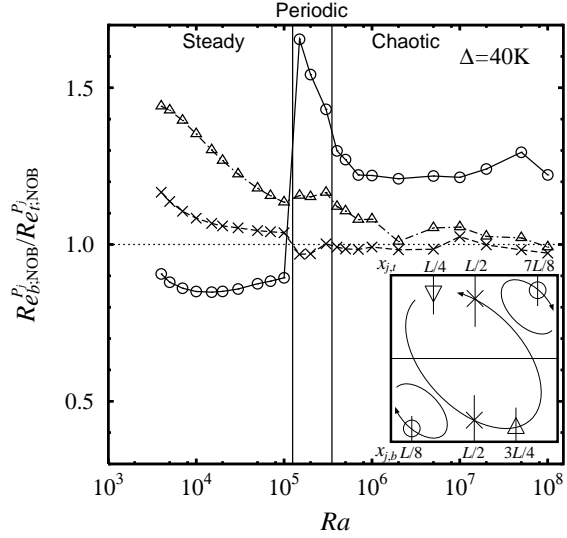,width=8cm,angle=0}
\end{center}
\caption{
Ratio $Re^{P_j}_{b;NOB}/Re^{P_j}_{t;NOB}$ vs.\ $Ra$ for the different peak positions in the cell. Note 
that due to the center-point symmetry of the conditionally time averaged field $ \overline{\bm u}$, the peak 
positions along the lines $x_{j,b} = \tfrac{L}{8}, \tfrac{L}{2}, \tfrac{3L}{4}$
for the bottom part have to be compared with the peak positions along $x_{j,t} = \tfrac{7L}{8}, \tfrac{L}{2}, 
\tfrac{L}{4}$ for the top part. The figure shows that the main convection loop for large $Ra$ establishes 
a uniform velocity amplitude, while the secondary counter-rotating rolls do not enjoy the same property. Instead, 
since they are BL-dominated rather than bulk-dominated, they show significant NOB bottom-top asymmetry. Note that 
the accuracy of this test may be assessed from 
the level of center-point asymmetry we had in the OB case, which was always below 5 $\%$. Therefore we may 
conclude that in the chaotic regime there are no NOB deviations distinguishable for the 
$Re^{P_j}_{b;NOB}/Re^{P_j}_{t;NOB}$ ratios at the two positions $x_{2,3} = \tfrac{L}{2}, \tfrac{3L}{4}$. But 
there is a significant deviation for $L/8$, i.e., for the secondary rolls' ratio.
} 
\label{onewind}
\end{figure}

\begin{figure}

\vspace*{0.5cm}

\begin{center}
\vspace*{-2em}
\epsfig{file=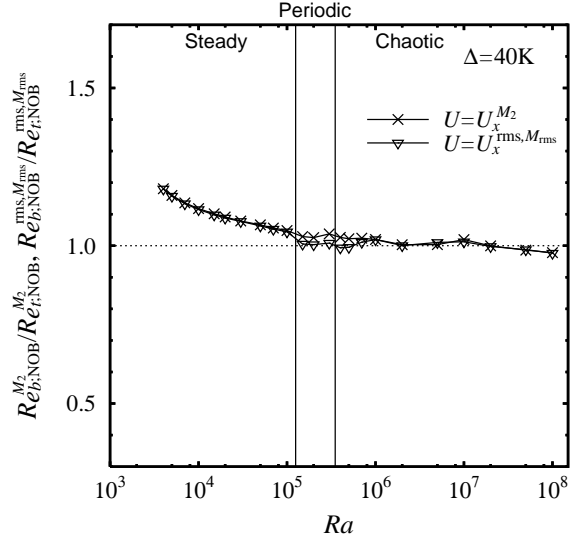,width=7.3cm,angle=-90}
\end{center}

\caption{
Same as Figure \ref{onewind}, but with the peak values of the horizontal area averaged velocities $U_x^{M_2}$
and $U_x^{rms,M_{rms}}$ taken for the velocity scales.
} 
\label{onewind_u_urms}
\end{figure}

We now study the NOB case, starting with the conditionally time averaged velocity field $\overline{\bm u}$, which 
now is bottom-top asymmetric. In particular, there is thinning/thickening of the 
bottom/top kinetic BLs. For the wind amplitudes based on the peak values of $\overline{\bm u}$ along vertical lines 
we can distinguish between bottom and top peak velocities $U^{P_j}_b$ and $U^{P_j}_t$ for the various $x_j$-lines. 
As shown in Fig.\ \ref{onewind}, the ratio $Re^{P_{x=L/2}}_b~/~Re^{P_{x=L/2}}_t$ characteristic for the peak 
velocities taken in the main primary circulation roll converges to unity as $Ra$ is increased up to $Ra \approx 10^8$
and $\Delta$ kept fixed. This shows that for the primary roll, which scans the bulk of the convection cell, a single 
velocity amplitude develops in the turbulent range also under NOB conditions, just as assumed in the BL theory 
introduced in \cite{ahl06,ahl07}. This is also found in Figure \ref{onewind_u_urms}, where the peak 
velocity scales are replace by $U_x^{M_{1,2}}$ and $U_x^{rms,M_{rms}}$. In particular, the large variation in the 
bottom/top Reynolds number ratio in Figure \ref{onewind}, which reflects the spatial inhomogeneity in the velocity 
distribution, is smoothened by taking the area average as indicated by the almost flat profiles of $U_x^{M_i}$ and 
$U_x^{rms,M_{rms}}$ in Figure \ref{onewind_u_urms}.

The secondary rolls, counter-rotating to the primary roll, have different wind amplitudes near bottom and top, as 
becomes apparent in the bottom-top ratio at the line $x_{1,b}/L = 0.125$ and its mirror at the top 
$x_{1,t}/L = 1 - 0.125$. The secondary roll at the warmer bottom plate is faster than the corresponding 
secondary roll near the top plate. We understand this from the smaller viscosity near the bottom due to the higher 
temperature. For the working fluid water and $\Delta = 40 $K the bottom-top asymmetry is as large as $\approx 25$\%   

\begin{figure}
\vspace*{0.3cm}
\ \hspace{5em}\ (a)
\begin{center}
\vspace*{-2em}
\epsfig{file=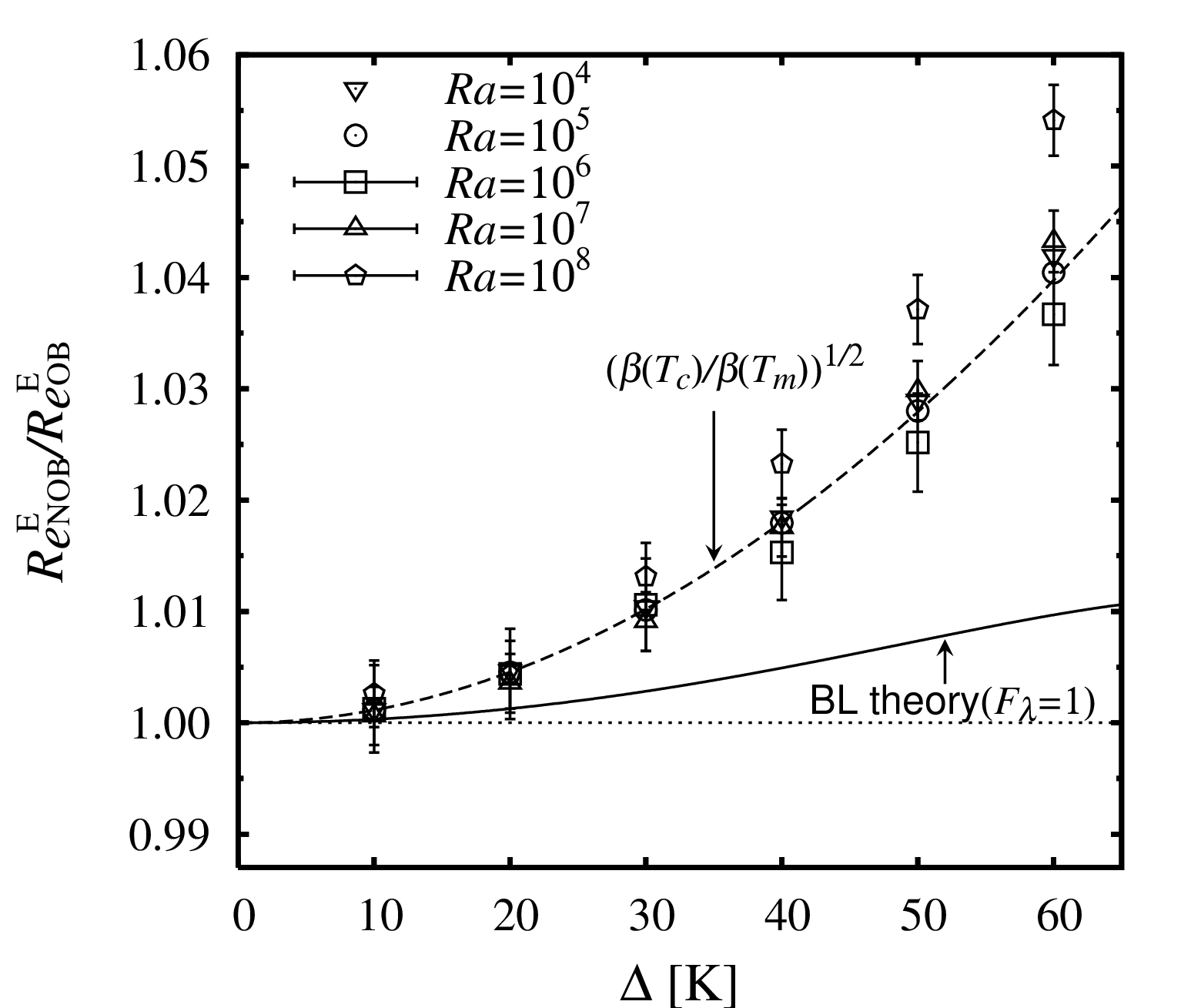,width=7cm,angle=-90}
\end{center}

\vspace*{0.3cm}

\ \hspace{5em}\ (b)
\begin{center}
\vspace*{-2em}
\epsfig{file=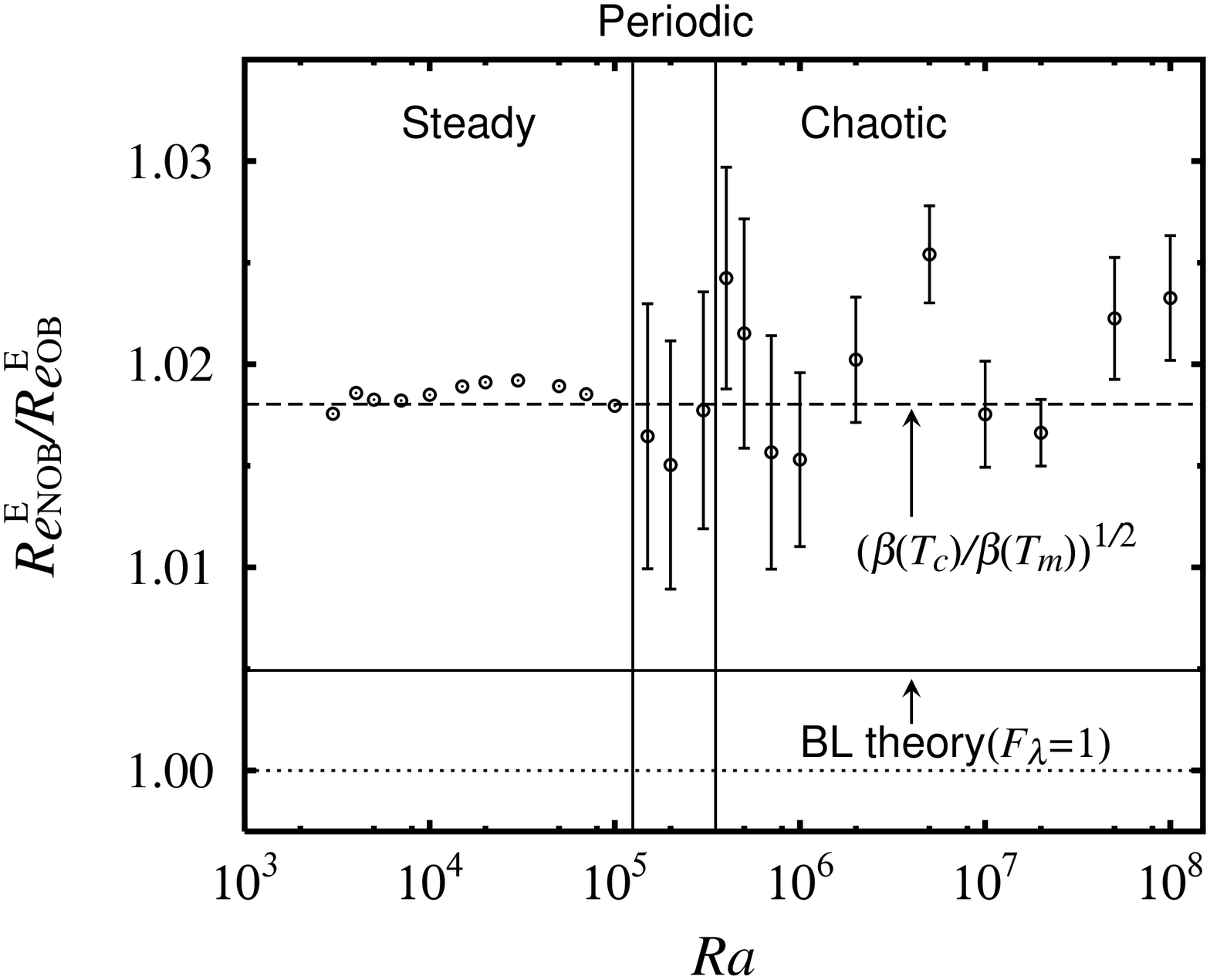,width=7cm,angle=-90}
\end{center}
\caption{
The Reynolds number ratio $Re^E_{NOB}/Re^E_{OB}$ based on the total kinetic energy for water at $T_m=40^o$C. 
Upper: $Re^E_{NOB}/Re^E_{OB}$ versus $\Delta$ for various choices of fixed $Ra$.
Lower: $Re^E_{NOB}/Re^E_{OB}$ versus $Ra$ at fixed non-Oberbeck-Boussinesqness $\Delta = 40$K.
The full line is calculated from BL theory under the assumption that $\lambda_b^{sl} + \lambda_t^{sl} 
= 2 \lambda_{OB}^{sl}$, i.e., $F_{\lambda} = 1$. The dashed line indicates the limit $\Delta \rightarrow 0$ where 
the Reynolds number ratio approaches $1$. For the explanation in terms of the $T$-dependence of $\beta$ see text, in 
particular (\ref{eq:ree_ratio_bet}). 
} 
\label{ratio_rev}
\end{figure}

We therefore calculated the NOB/OB ratio of $Re^E$ versus $\Delta$ for various values of $Ra$ as well as versus $Ra$ 
at fixed non-Boussineqness $\Delta=40 $K. This is shown in Fig.~\ref{ratio_rev}.
We observe that NOB effects are clearly present, although rather weak only, of order $2\%$, indicating 
a small increase of the kinetic energy based mean velocity. The reason for the rather small NOB effect on 
the global wind amplitude in spite of the large changes of the bottom and top velocities is that the 
secondary rolls only contribute a limited fraction to the global volume average. Also, the changes of the 
velocities of the secondary rolls are opposite in sign, one contributing a larger (bottom) the other contributing 
a smaller (top) amplitude. The remaining net change of the global amplitude $U^E$ thus again is due to the nonlinear 
temperature dependences of the material parameters, producing different secondary roll velocities at bottom and top. 
This crucial importance of the nonlinearities in the temperature dependence of the material properties was already 
found in \cite{ahl06}.   

\begin{figure}
\vspace*{0.3cm}
\ \hspace{5em}\ (a)
\begin{center}
\vspace*{-2em}
\epsfig{file=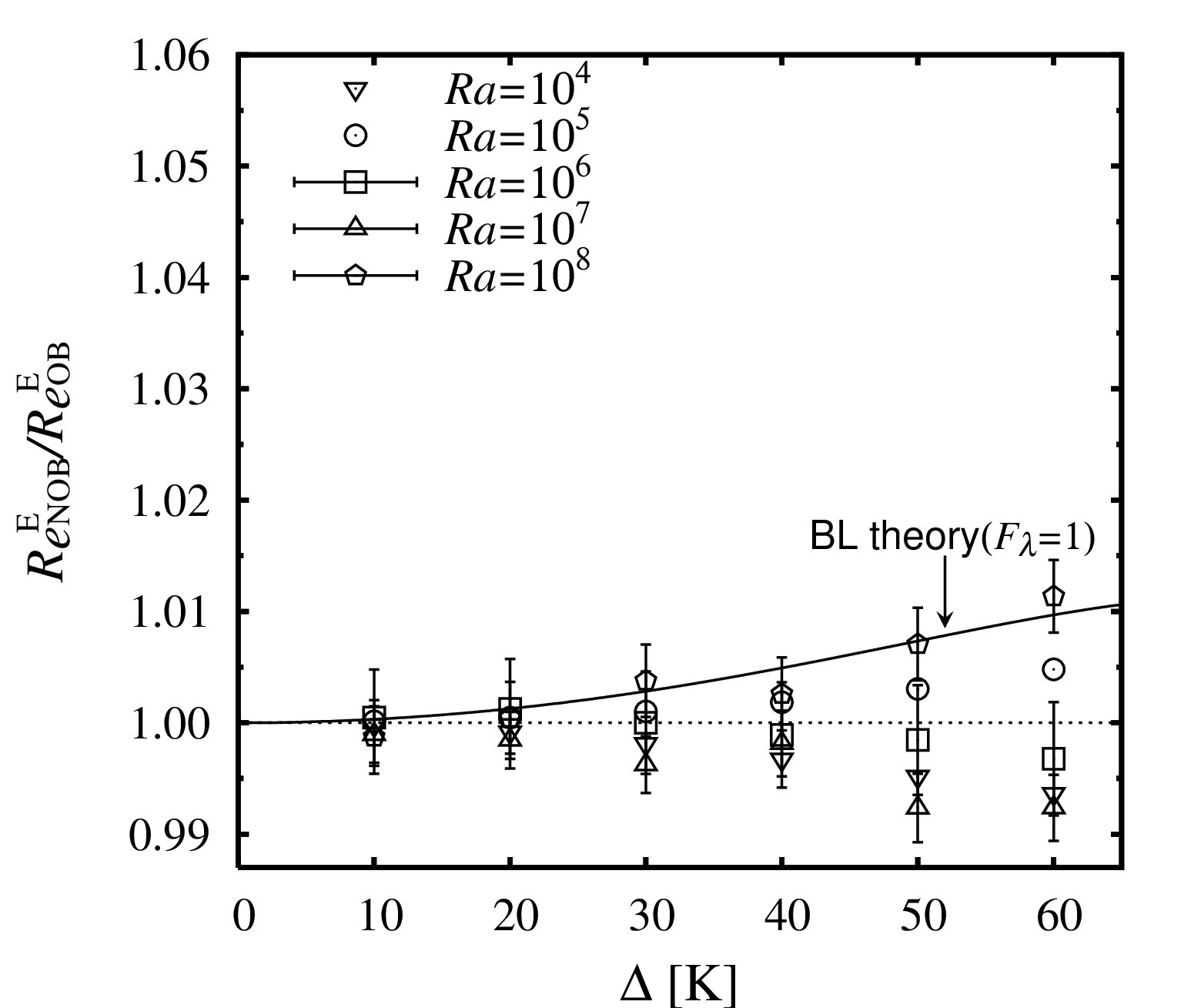,width=7cm,angle=-90}
\end{center}

\vspace*{0.3cm}

\ \hspace{5em}\ (b)
\begin{center}
\vspace*{-2em}
\epsfig{file=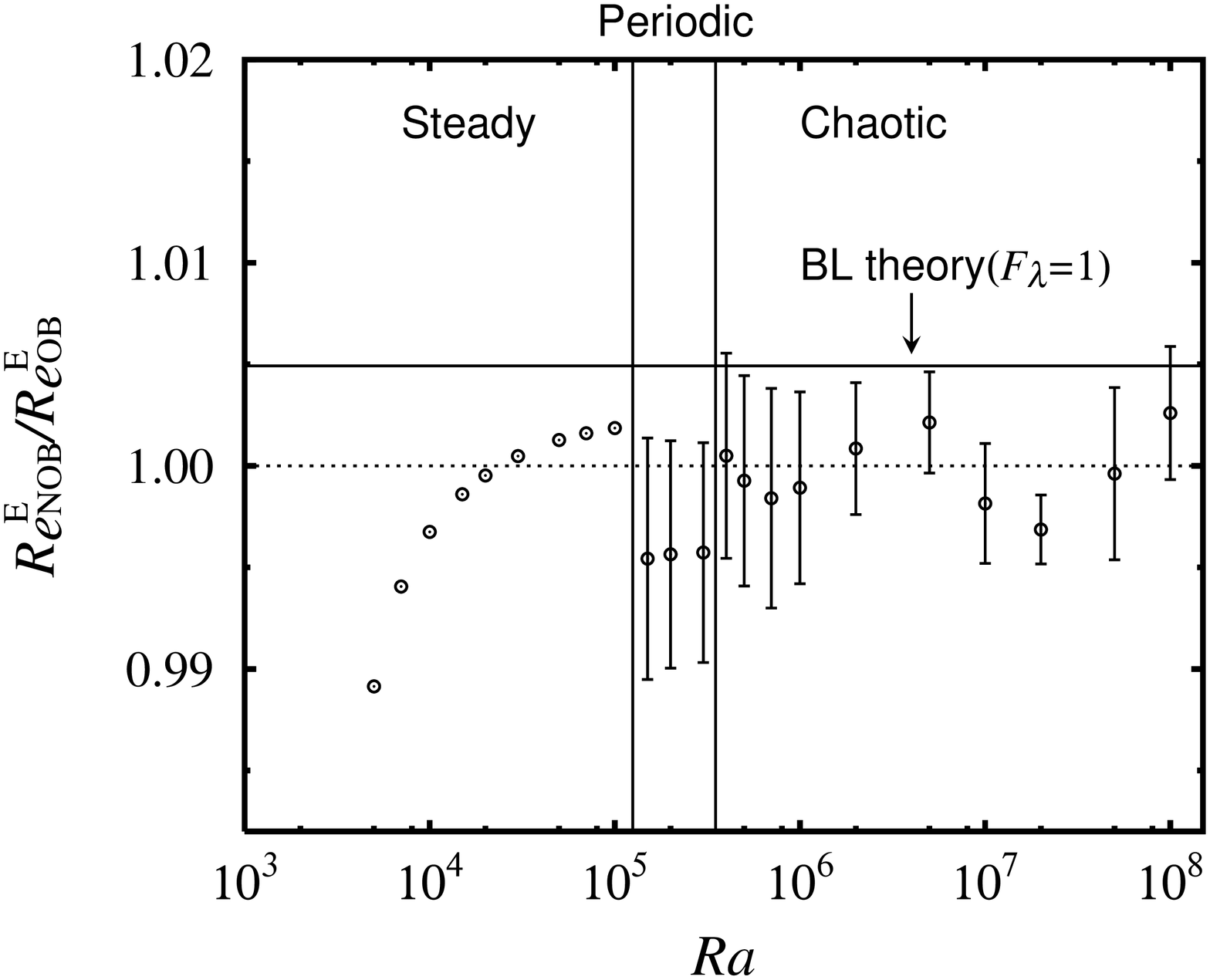,width=7cm,angle=-90}
\end{center}

\caption{Same as Figure \ref{ratio_rev}, upper, but here with only linear dependence of $\rho(T)$ on the  
temperature $T$, meaning that the thermal expansion coefficient $\beta(T)$ is set to the constant value $\beta_m$.
Note that under this approximation there is good agreement with the results from BL theory and $F_{\lambda} = 1$
for large $Ra$, which means that the thermal convection takes significant notice of the full $T$-dependence 
of the thermal expansion coefficient $\beta$. BL theory misses that per construction.
}
\label{ratio_rev_betac}
\end{figure}

\begin{figure}
\vspace*{0.3cm}
\ \hspace{5em}\ (a)
\vspace*{-2em}
\begin{center}
\epsfig{file=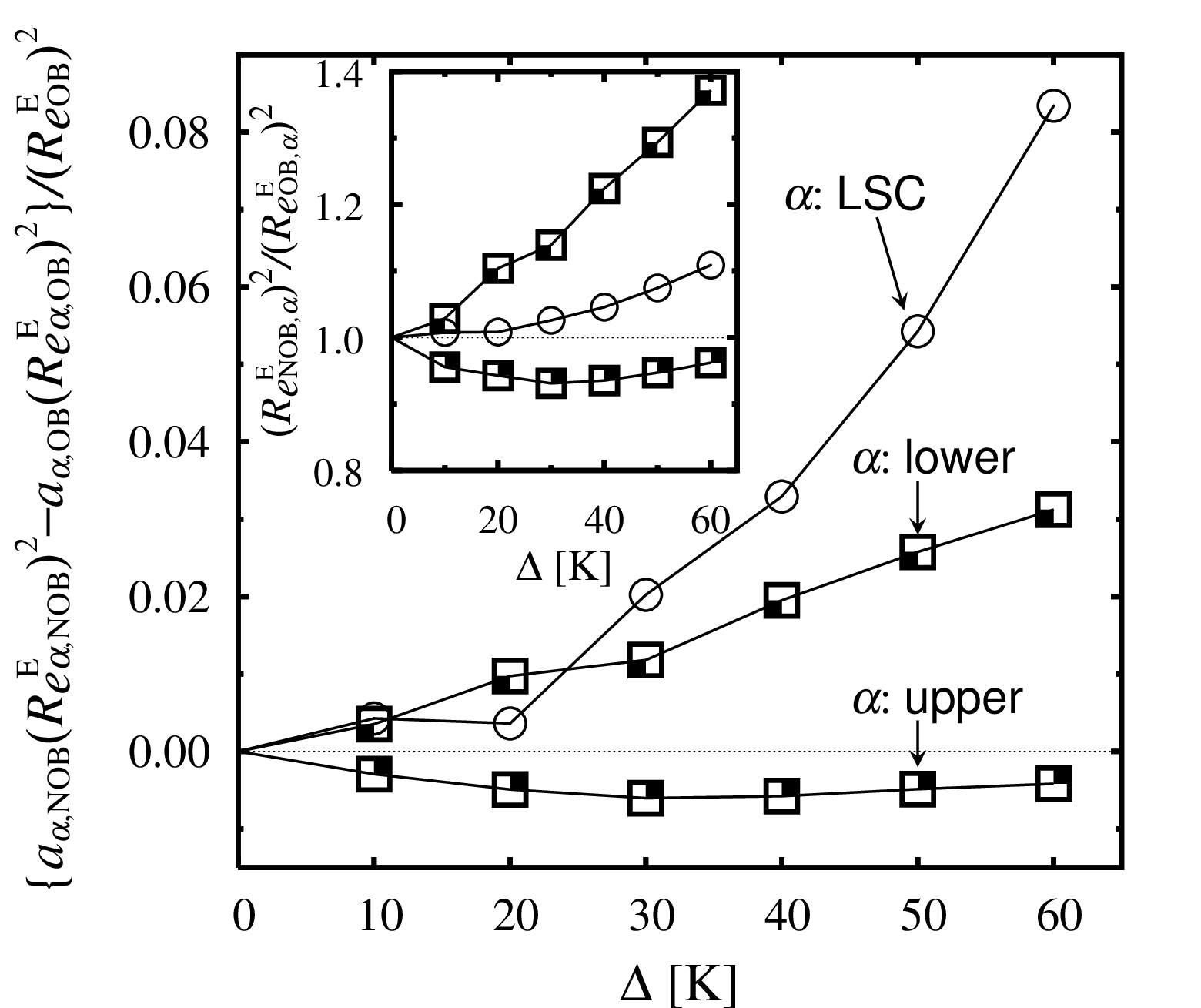,width=6.5cm,angle=-90}
\end{center}
\vspace*{0.3cm}
\ \hspace{5em}\ (b)
\begin{center}
\vspace*{-2em}
\epsfig{file=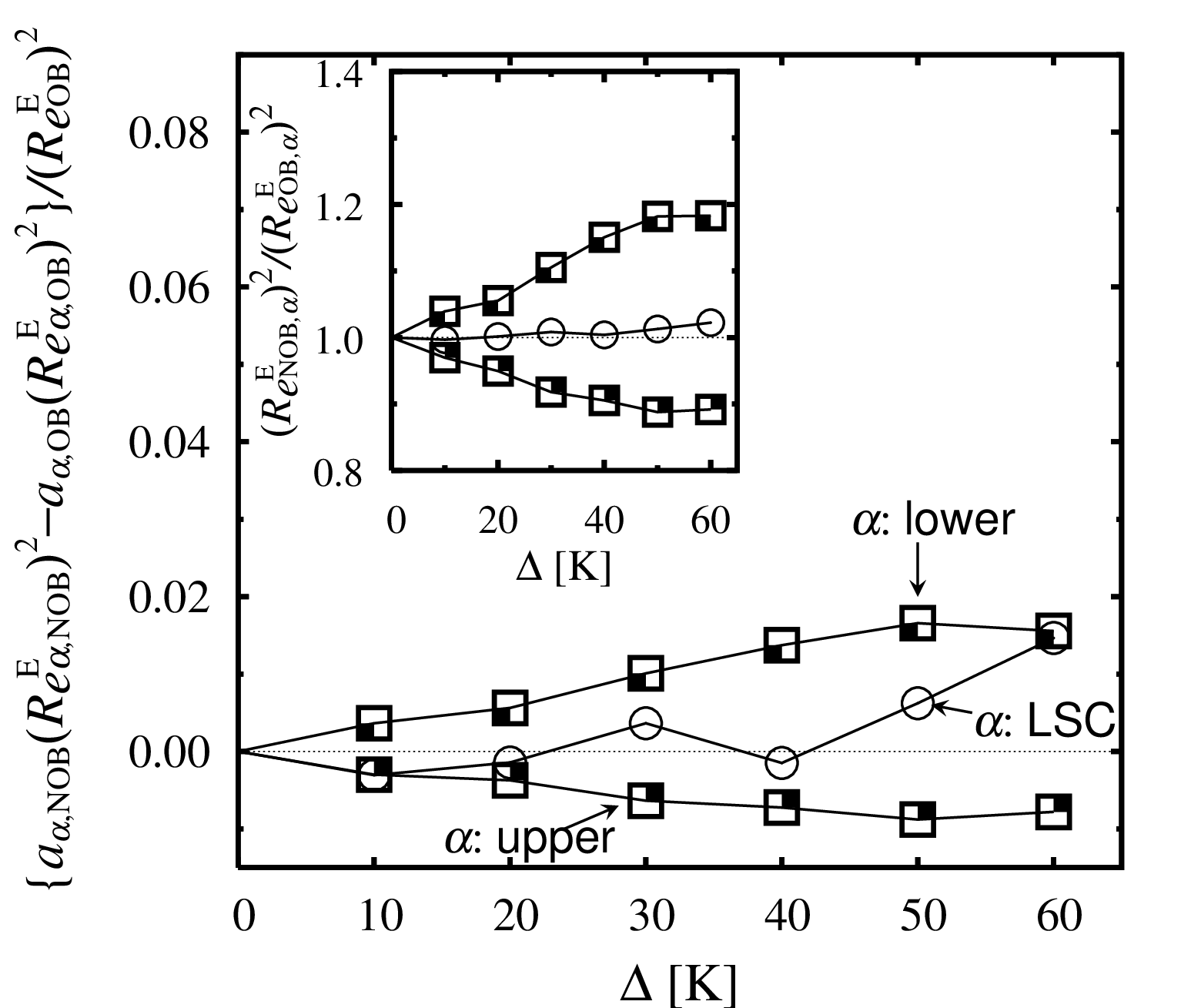,width=6.5cm,angle=-90}
\vspace*{0.3cm}
\end{center}
\caption{ 
The relative deviation of the kinetic energy 
$\{a_{\alpha, NOB}(Re^E_{\alpha,NOB})^2 - a_{\alpha,OB}(Re^E_{\alpha,OB})^2\}$ $/(Re^E_{OB})^2$
in region $\alpha$ due to NOB effects for water at $Ra=10^8$ and $T_m=40^o$C. The label $\alpha$ 
denotes the portion of the large scale circulation (LSC) roll in the center region 
(as indicated by the solid lines in Figure 
\ref{snap_cs}), or the secondary counter-rotating rolls in the lower and upper corners (given by the dashed lines in 
Figure \ref{snap_cs}). The partition is determined by using the sign of $\psi=\int_0^z{\rm d}\hat{z}\ 
\overline{u_x}^c(x,\hat{z})$, e.g., the lower corner flow is defined as the region satisfying 
$\psi\leq 0$, $x\leq L/2$ and $z\leq L/2$. Upper panel: the full temperature dependence is considered for the 
buoyancy $g(1-\rho/\rho_m)$ as given in Table \ref{tab1}. Lower panel: restriction to only linear 
dependence of the buoyancy with 
respect to the temperature $T$, i.e., the thermal expansion coefficient $\beta(T)= \beta_m$ is constant. 
The insets show the (squared) NOB/OB Reynolds number ratio in the each region.
}
\label{dev_rev2}
\end{figure}

Let us recall that the Nusselt number becomes smaller with increasing $\Delta$ as was shown in Fig.~\ref{f-vs-del}.
Therefore, although the global Reynolds number ratio is enhanced under NOB conditions,
the overall heat transport is attenuated. We have also included in Fig.~\ref{ratio_rev} a comparison with BL theory. 
We remind that to make this theory predictive an additional assumption on $F_{\lambda}$ (e.g. $F_{\lambda}\simeq 1$)
has to be made. Note that the extended BL theory in this form under-estimates the NOB effect on the wind.

As the origin of this discrepancy we can now identify the effect of the temperature dependence of the thermal expansion 
coefficient, which is not included in the BL equations in  \cite{ahl06}. For this explanation we offer the 
following argument. As a naive estimate one can assume that the volume averaged velocity scale $U^E$
in essence should coincide with the free-fall velocity, that is $U^E \simeq \sqrt{g L \beta(T) \Delta}$. 
This gives the scaling $Re \sim Pr^{-1/2}Ra^{1/2}$. Assuming then that the temperature 
of the bulk is dominated by $T_m$ and $T_c$ respectively in the OB and NOB cases, one gets  
\begin{equation}
\frac{Re^E_{NOB}}{Re^E_{OB}} \sim 
\left(\frac{\beta(T_c)}{\beta(T_m)}\right)^{1/2}.
\label{eq:ree_ratio_bet}
\end{equation}
This is in encouraging agreement with the DNS data, cf. Figure \ref{ratio_rev}.  -- As a further 
support of our argument emphasizing 
the importance of the temperature dependence of the  thermal expansion coefficient in the bulk 
we also calculated $Re^E_{NOB}/Re^E_{OB}$ for a hypothetical liquid, which has all material properties as  
water, apart from the thermal expansion coefficient $\beta$, which we keep constant at $\beta_m$, 
see Fig.~\ref{ratio_rev_betac}. Indeed, $Re^E_{NOB}$ now only shows a smaller than 1\% deviation from
$Re^E_{OB}$, even at $\Delta = 60$K. These tiny deviations from $Re^E_{NOB}/Re^E_{OB}= 1$ are consistent
with the results from the Prandtl Blasius theory with the additional assumption $F_\lambda = 1$. 
Fig.~\ref{ratio_rev_betac} thus confirms that the main origin of the NOB deviation in the Reynolds number $Re^E$ 
is the temperature dependence of the thermal expansion coefficient, an effect which clearly cannot
be captured in the Prandtl-Blasius BL theory.

The changes of $Re^E$ caused by the loss of OB conditions can be analysed quantitatively in detail.
For this we decompose the volume average into three regions, corresponding to the main, primary, large scale 
circulation (LSC) and the two secondary lower and upper corner rolls.  
\begin{equation}
(Re^E)^2=a_{LSC}(Re_{LSC}^{E})^2+a_{lower}(Re_{lower}^{E})^2+a_{upper}(Re_{upper}^{E})^2.
\label{eq:decomp_ree2}
\end{equation}
Here $a_\alpha$ denotes the ratio of the volume $V_\alpha$ occupied by the region $\alpha$, with 
$\alpha$ = LSC, lower, or upper secondary roll, to the total volume, and 
$Re_{\alpha}^E$ is the Reynolds number based on 
the kinetic energy averaged over each volume $V_\alpha$, 
\begin{equation}\label{re_alpha}
Re_{\alpha}^{E} = \frac{LU_\alpha^E}{\nu_m} ~~~\mbox{with} ~~~ U_\alpha^E = \sqrt{\frac{1}{V_\alpha}
\int_{V_{\alpha}}\!\!\!\!\!{\rm d}^3{\bm x}\ \frac{1}{2}\left(\overline{u_x^2}({\bm x}) + 
\overline{u_z^2}({\bm x})\right)}.
\end{equation}
The lower secondary roll is characterized by a clockwise rotation and negative values of the 
$\overline{\bm u}(\bm x)$-stream function $\psi(x,z)=\int_0^z{\rm d}\hat{z}\ \overline{u_x}(x,\hat{z})$.
The region $V_{lower}$ of the lower corner roll is defined by $\psi\leq 0$, $x\leq L/2$ and $z\leq L/2$, while for 
the upper corner roll we have $\psi\leq 0$, $x\geq L/2$ and $z\geq L/2$. The remaining region comprises the primary main
roll, LSC. 

The insets of Figure \ref{dev_rev2} show the (squared) NOB/OB Reynolds number ratios 
$(Re_{\alpha,NOB}^{E})^2$ $/(Re_{\alpha,OB}^{E})^2$ for each region $\alpha$ at $Ra=10^8$.
The insets both in the upper and lower panels reveal that the largest NOB enhancement of the kinetic energy occurs 
inside the lower corner secondary roll (near the warmer bottom plate), as already found in the bottom-top asymmetry 
of the peak velocity $Re^{P_{x=L/8}}_{b,NOB} / Re^{P_{x=7L/8}}_{t,NOB}$ in Figure \ref{onewind}. But even this 
enhancement does not impact significantly on the overall $Re^E$-change because the volume ratio of each corner flow 
is only $a_{lower} = a_{upper} \approx 9$\% while the main volume fraction is $a_{LSC} \approx 82$\%. 
This is even more pronounced in the OB case, where $a_{lower}(Re_{lower}^E)^2/(Re^E)^2 = 0.05$ and
$a_{LSC}(Re_{LSC}^{E})^2 /(Re^E)^2 = 0.90$, while for the NOB case at $\Delta=40$K these fractions are $0.07$ and
$0.04$ for the lower and upper secondary rolls, and the primary LSC contributes $0.89$. 
To visualize these contributions of each subvolume $\alpha$ to the overall change, we show in 
Figure \ref{dev_rev2} the normalized kinetic energy deviation 
$\{a_{NOB,\alpha}(Re^{E}_{NOB,\alpha})^2 - a_{OB,\alpha}(Re^{E}_{OB, \alpha})^2\}/(Re^{E}_{OB})^2$ 
due to the NOB effect. The upper panel proves that the NOB enhancement of the total kinetic energy 
is primarily due to the LSC and secondarily to the lower corner roll. For comparison we plotted the corresponding 
deviations in the lower panel for a hypothetical fluid with thermal expansion coefficient fixed at $\beta_m$. Then
the enhancement of the LSC contribution is much smaller, while the attenuation and enhancement respectively 
in the upper and lower secondary rolls are comparable and in addition compensate each other. This leads to the 
much smaller change in the total kinetic energy in Figure \ref{ratio_rev_betac} as compared with allowing the full 
temperature dependence of $\beta(T)$ in Figure \ref{ratio_rev}. Apparently the {\em nonlinear} temperature dependence 
of the buoyancy is very important. Note that the derivative of the buoyancy 
$\partial \left( g(1-\rho/\rho_m) \right) / \partial T$, corresponding to the driving force per temperature 
displacement, increases when increasing the temperature deviation $(T-T_m)$ because the coefficient $C_2$ for the 
buoyancy expression as reported in Table \ref{tab1} is positive. Therefore the buoyancy force gets larger and 
the bulk kinetic energy is more enhanced for given temperature deviation $T-T_c$, 
if the mean bulk temperature $T_c$ is larger than $T_m$ as observed in Figures \ref{tc-vs-del} and \ref{tc-vs-ra}.

\section{Summary and conclusions} \label{sec6}

In summary, we have studied the temperature profile, the heat current density, and the properties of the 
large scale convections as defined by several representative velocity scales. The center temperature 
$T_c$ and the Nusselt number ratio $Nu_{NOB}/Nu_{OB}$ resulting from the two-dimensional numerical NOB 
simulations are in good agreement with the available experimental data for water (\cite{ahl06}).  
\cite{ahl06}'s experimental finding $F_{\lambda} = 1$ for water is argued to be incidental, originating from the 
specific temperature dependence of the material constants of water at $40^o$C. This finding cannot be 
generalized to other fluids or to other mean temperatures. 
For water the heat flux reduction due to the deviations from OB conditions is for all practical purposes 
due to the modified temperature drops over the BLs, represented by $F_{\Delta}$, whereas for other working 
fluids it is influenced also by the changes of thermal BL thicknesses, expressed by $F_{\lambda}$.

The results of the simulations also agree with \cite{ahl06}'s predictions for the 
central temperature, which is based on an extended Prandtl-Blasius theory.
As that theory ignores plumes and side wall effects, these apparently hardly contribute 
to the determination of the central temperature $T_c$. However, physically they
do contribute to the shape of the temperature profiles. Our simulations
reveal their increasing effect on the profiles with increasing $Ra$. 
As the overall heat transfer is determined by the slope of the temperature profiles at the plates, 
it is to be expected that with increasing $Ra$ the plumes increasingly affect the Nusselt
number, in coherence with \cite{gro00,gro01,gro02,gro04}'s unifying theory.

The second part of the paper is devoted to the flow organization in the OB and the NOB cases.
First of all, also in the NOB case the large scale convection roll is characterized by only one
velocity scale. In contrast, the top and bottom corner flows have different velocity scales in the NOB case,
reflecting the enhanced and reduced viscosities close to the respective plates. We defined various
different velocity scales, based on global and area averages and peaks in the profiles and analyzed
how these change under NOB conditions. For the ratio of the energy based Reynolds numbers 
which also is representative for the others 
we find 
$Re_{NOB}^E/Re_{OB}^E \approx (\beta(T_c)/\beta(T_m))^{1/2}$, i.e., NOB deviations in the
Reynolds number are strongly 
governed by the temperature dependence of the thermal expansion coefficient.
This finding suggests that fluids which display no or only a weak temperature dependence of the
thermal expansion coefficient hardly show any NOB effects on the Reynolds numbers. 

From our point of view the two next steps in numerical work on NOB correction for RB flow are:
(i) Study the detailed modifications of the various BL thicknesses and profiles through NOB effects, 
and (ii) confirm that at least for Prandtl number of 1 and larger the findings of this paper also hold for
 three-dimensional
RB flow. Moreover, NOB  experiments focusing on the flow organization, BL layers, and Reynolds number
modifications would be very desirable.

\begin{acknowledgments}
\noindent
{\it Acknowledgment:} We thank Guenter Ahlers and Francisco Fontanele Araujo for many fruitful discussions over the last years. The work in Twente is part of the research program of FOM, which is financially supported by NWO. S.Gn acknowledges support by FOM, E.C. acknowledges support by CNRS.
\end{acknowledgments}


\end{document}